\documentclass[useAMS,usenatbib]{mn2e}
\usepackage{graphicx}
\usepackage{color}
\usepackage{amssymb}

\title[Optical and radio properties of giant radio quasars: Central black hole characteristics.]{Optical and radio properties of giant radio quasars: Central black hole characteristics.}
\author[A. Ku\'zmicz and M. Jamrozy]{A. Ku\'zmicz$^{1}$\thanks{E-mail: cygnus@oa.uj.edu.pl} and M. Jamrozy$^{1}$\thanks{E-mail: jamrozy@oa.uj.edu.pl}\\
$^{1}$Astronomical Observatory, Jagiellonian University, ul. Orla 171 , 30-244 Krak\'ow, Poland}
\begin{document}

\date{Accepted 2011 Month 00. Received 2011 Month 00; in original form 2011 Month 00}

\pagerange{\pageref{firstpage}--\pageref{lastpage}} \pubyear{0000}

\maketitle

\label{firstpage}

\begin{abstract}
We analysed the optical and radio properties of lobe-dominated giant-sized ($>$ 0.72 Mpc) radio quasars and compared the results with those derived for a sample of smaller radio sources to determine whether the large size of some extragalactic radio sources is related to the properties of their nuclei. We compiled the largest (to date) sample of giant radio quasars, including 24 new and 21 previously-known objects, and calculated a number of important parameters of their nuclei such as the black hole mass and the accretion rate. We conclude that giant radio quasars have properties similar to those of smaller size and that giant quasars do not have more powerful central engines than other radio quasars. The results obtained are consistent with evolutionary models of extragalactic radio sources which predict that giant radio quasars could be more evolved (aged) sources compared to smaller radio quasars. In addition we found out that the environment may play only a minor role in formation of large-scale radio structures.
\end{abstract}
 
\begin{keywords} 
galaxies: active - quasars: emission lines - radio continuum: galaxies
\end{keywords}

\section{Introduction}
Giant radio sources (GRSs) are defined as powerful extragalactic radio sources, hosted by galaxies or quasars, for which the projected linear size of their radio structure is larger than 0.72 Mpc{\footnote {Many authors, assuming $H_0=50$ km s$^{-1}$Mpc$^{-1}$, have used 1 Mpc as the defining size for GRSs. For the currently accepted cosmological parameters as given above, this size decreases to $\sim$0.72 Mpc.}} (assuming $H_0=71$ km s$^{-1}$Mpc$^{-1}$, $\Omega_M=0.27$, $\Omega_{\lambda}=0.73$; \citealt{b59}). Looking through the new, ``all-sky'' radio surveys such as the Westerbork Northern Sky Survey (\citealt{b51}), the NRAO VLA Sky Survey (NVSS; \citealt{b13}), the Faint Images of the Radio Sky at Twenty centimeters (FIRST; \citealt{b3}), the Sydney University Molonglo Sky Survey (\citealt{b10}) and the Seventh Cambridge Survey (\citealt{b41}) a large number of new giant sources were identified. Almost all of these GRSs are included in the samples of giants presented by \cite{b14}, \cite{b34}, \cite{b36}, \cite{b37}, \cite{b54}, \cite{b56}, as well as in the list of giants known before 2000 published by \cite{b25}. GRSs are very useful in studying a number of astrophysical problems, for example the evolution of radio sources, the properties of the intergalactic medium (IGM) at different redshifts, and the nature of the central active galactic nuclei (AGN). It is still unclear why such a small fraction of radio sources reach such a large size -- it may be due to special external conditions, such as lower IGM density, or due to the internal properties of the ``central engine''. Our knowledge about the nature of GRSs has improved somewhat following studies conducted in the last decade. However, these were focused almost exclusively on: the role of the properties of the IGM (\citealt{b1111, b32}), the advanced age of the radio structure (e.g. \citealt{b1112, b37b}), and recurrent radio activity (e.g. \citealt{b1114, b1130}) as responsible for gigantic size.

To date, there are about 230 GRSs known, and just a small fraction of them ($\sim 8$ per cent) are actually related to quasars. The lobe-dominated radio quasars usually have a classical FRII (\citealt{b19}) morphology and most of their radio emission originates in the extended regions with steep radio spectra. The ratio of the flux density of the core to that of the lobes at 5 GHz is usually less than 1 (\citealt{b24}). Optically, the lobe-dominated radio quasars are similar to the core-dominated quasars (\citealt{b2}) and their lobe dominance is just an orientation effect.

In general, it is believed that strong jet activity in an AGN is related to the parsec or sub-parsec scale condition of its host galaxy and specifically to the properties of its central black hole (BH; \citealt{b6, b8, b7}). Furthermore, if assuming that the power of the ``central engine'' is responsible for the linear size evolution of an radio source, we should expect the largest objects to be related to radio quasars, which host the most energetic AGNs, as opposed to radio galaxies. It is because there is observational evidence for a correlation between jet power and the expansion speed of a radio source's lobes (e.g. \citealt{b1119, b1120}). The jets of high-power sources carry a larger momentum flux, which in turn implies a greater flux of kinetic energy. Therefore, radio quasars, which are on average more luminous than radio galaxies, should have higher lobe expansion speeds than radio galaxies and, assuming similar mean lifetimes of both these types of AGNs, radio quasars have the potential to reach larger size. A typical lobe expansion speed could be of the order of a few hundredths (or more) of the speed of light (e.g. \citealt{b1121}), and this, together with the typical lifetime of an evolved radio source of the order of a few times $10^7$ years, gives a size of the order of a Mpc.

The radio loudness of quasars still remains a debated issue. Radio observations of optically-selected samples of quasars showed that only 10-40 per cent of the objects are powerful radio sources (for reference see e.g. \citealt{b1122, b1123}). Recently, thanks to FIRST -- the large-area radio survey -- the number of quasars with faint radio fluxes has grown enormously. Therefore, it is now possible to investigate the optical and radio properties of quasars based on statistically large samples of objects (e.g. \citealt{b69, b1128, b1124, b1122, b1117, b1123}), and to try to understand the connection between the optical emission (luminosity, BH mass and spin, accretion rate) and the radio (jet) activity. Evidence that the spin of the BH plays a significant role in radio activity has recently been found (e.g. \citealt{b1115, b1116, b1117}). The relation between BH mass and radio loudness has also been intensively studied, but so far the results are equivocal. Many authors (e.g. \citealt{b35, b42, b17, b39, b44}) have found that, on average, radio-louder AGNs possess larger BH masses. However, there are also many reports arguing against any dependence between these quantities (e.g., \citealt{b48, b23, b71, b1122, b58}). Furthermore, the importance of the mechanical energy of jets and lobes released by BHs and the feedback on the surroundings has recently been realized (\citealt{b1125, b1127, b1126}). AGNs deposit large amounts of energy into their galactic environment which may, for example, be responsible for halting star formation. There is broad observational evidence that mechanical heating by jets plays an important role in balancing radiative losses from the intra-cluster medium. Radio jets and lobes of quasars can modify the structure of the environment not only on kpc scales but also, by the giant-size sources, on Mpc scales.
       
The aim of this study is to investigate the radio and optical properties for a sample of lobe-dominated giant radio quasars (GRQs). We would like to answer the question whether the size of GRQs is related to the internal properties of their hosts. To investigate the role and importance of the central engine in generation of Mpc-scale structures we have compiled the largest sample of GRQs to date.  

The paper is organized as follows: in Sect.~2 we describe our source samples and in Sect.~3 the possible biases of the samples. In Sect.~4 we present definitions of the parameters used in the analysis; in Sect.~5 we investigate the relations between the optical and radio properties of our sources. Sect.~6 presents our conclusions.\\ 
Throughout the paper we assume the standard cosmology, with parameters as provided at the beginning of this Section.

\section[]{The sample}
In our analysis we use 45 GRQs of which 21 are taken from the existing literature (for details see Table 1). The remaining 24, which were not previously identified as GRQs, we selected from catalogues of radio quasars compiled by \cite{b20}, \cite{b4}, \cite{b69}, and \cite{b65}. The presented sample of giant-sized radio quasars is the largest to date. It contains sources even at large redshifts ($z\sim 2$).

As a comparison sample, we selected 31 smaller lobe-dominated radio quasars from a list of radio sources given by \cite{b47}. In order to obtain a number of objects comparable to the GRQ sample, 18 quasars selected from the catalogues cited were added to the comparison sample. The linear sizes of these objects are close to the limiting size of 0.72 Mpc, as we wanted to have a smooth transition in linear size between the smaller radio quasars and the GRQs. The sources from the comparison sample of lobe-dominated radio quasar meet
the following criteria:

\begin{enumerate}
\item Have optical spectra in the Sloan Digital Sky Survey (SDSS; \citealt{b1}).

\item Possess the MgII(2798\AA) broad emission line in their spectra (as most of our GRQ spectra contain the MgII(2798\AA) line). This condition limits the range of the redshift to $0.4\lesssim z \lesssim 2$;  it was adopted in order to have similar properties of the optical spectra for all quasars and hence allow homogeneous measurements using the same methods for both samples.

\item Have a projected angular size of the radio structure larger than 0\farcm2, to properly separate the components  (lobes and core) of the source in the FIRST maps (which have 5\arcsec$\times5$\arcsec\, angular resolution).
\end{enumerate}

\noindent
All our quasars possess a classical FRII radio morphology. These objects lie almost in the plane of the sky and therefore the influence of relativistic beaming is small. It is easy to determine the physical size of such sources based on radio maps, even at high redshift.

The final samples contain 45 GRQs and 49 smaller radio quasars, whose basic parameters are provided in Tables 1 and 2 respectively.  The new, previously unrecognised, GRQs are marked in bold type in Table 1. Optical spectra from the SDSS as well as radio maps from the NVSS and FIRST surveys are available for almost all of these objects. In addition, the spectra of nine quasars published by \cite{b69}, \cite{b4}, \cite{b65} and \cite{b21} were provided to us in electronic FITS format by R. White (these are marked by the letter W in Tables 1 and 2). The columns of Table 1 and 2 contain respectively: (1) - J2000.0 IAU name;  (2) and (3) J2000.0 right ascension and declination of the central position of the optical quasar;  (4) redshift of the host object; (5) angular size in arcmin; (6) projected linear size in Mpc; (7) availability of the spectrum from the SDSS survey (S), or provided by White (W); availability of radio maps from NVSS or FIRST (N or F, respectively); (8) references to the identified object. Unfortunately, for two GRQs,  J0631$-$5405 and J0810$-$6800, we had neither spectral nor radio data at hand, therefore we excluded them from further analysis.

\begin{table*}
 \centering
 \begin{minipage}{130mm}
  \caption{List of giant-sized ($>$ 0.72 Mpc) radio quasars.}
  \begin{tabular}{@{}lccccccl@{}}
  \hline
IAU                & \hspace{0.5cm}$\alpha$(J2000.0)& $\delta$(J2000.0)& z     &  d     & D      &  \hspace{0.5cm}Avail.& \hspace{0.5cm}Ref.\\
name               & \hspace{0.5cm}(h m s)       & ($^{o}$ \arcmin\, \arcsec)  &            &  arcmin& Mpc    &  \hspace{0.5cm}Data  &                  \\
(1)                & \hspace{0.5cm}(2)           & (3)             &(4)        &(5)     & (6)    &  \hspace{0.5cm}(7)  & \hspace{0.5cm}(8)\\
\hline
{\bf J0204$-$0944} & \hspace{0.5cm}02 04 48.29   & $-$09 44 09.5   &   1.004   & 6.035  & 2.914  &  \hspace{0.5cm}S,N,F& \hspace{0.5cm}1 \\
{\bf J0210$+$0118} & \hspace{0.5cm}02 10 08.26   & $+$01 18 42.3   &   0.870   & 2.618  & 1.214  &  \hspace{0.5cm}W,N,F& \hspace{0.5cm}1 \\ 
J0313$-$0631       & \hspace{0.5cm}03 13 32.88   & $-$06 31 58.0   &   0.389   & 3.090  & 0.973  &  \hspace{0.5cm}S,N  & \hspace{0.5cm}2 \\
J0439$-$2422       & \hspace{0.5cm}04 39 09.20   & $-$24 22 08.0   &   0.840   & 1.960  & 0.899  &  \hspace{0.5cm}N    & \hspace{0.5cm}3 \\
J0631$-$5405       & \hspace{0.5cm}06 32 01.00   & $-$54 04 58.7   &   0.204   & 5.200  & 1.04   &  \hspace{0.5cm}-    & \hspace{0.5cm}4 \\
J0750$+$6541       & \hspace{0.5cm}07 50 34.43   & $+$65 41 25.4   &   0.747   & 3.271  & 1.439  &  \hspace{0.5cm}N    & \hspace{0.5cm}5 \\
{\bf J0754$+$3033} & \hspace{0.5cm}07 54 48.86   & $+$30 33 55.0   &   0.796   & 3.842  & 1.730  &  \hspace{0.5cm}S,N,F& \hspace{0.5cm}6 \\
{\bf J0754$+$4316} & \hspace{0.5cm}07 54 07.96   & $+$43 16 10.6   &   0.347   & 8.061  & 2.360  &  \hspace{0.5cm}S,N,F& \hspace{0.5cm}7 \\   
{\bf J0801$+$4736} & \hspace{0.5cm}08 01 31.97   & $+$47 36 16.0   &   0.157   & 5.438  & 0.876  &  \hspace{0.5cm}S,N,F& \hspace{0.5cm}7 \\  
J0809$+$2912       & \hspace{0.5cm}08 09 06.22   & $+$29 12 35.6   &   1.481   & 2.184  & 1.118  &  \hspace{0.5cm}S,N,F& \hspace{0.5cm}6, 8 \\
                   &                             &                 &           &        &        &                     &                 \\
J0810$-$6800       & \hspace{0.5cm}08 10 55.10   & $-$68 00 07.7   &   0.231   & 6.500  & 1.42   &  \hspace{0.5cm}-    & \hspace{0.5cm}9 \\
J0812$+$3031       & \hspace{0.5cm}08 12 40.08   & $+$30 31 09.4   &   1.312   & 2.427  & 1.203  &  \hspace{0.5cm}S,N,F& \hspace{0.5cm}8 \\
J0819$+$0549       & \hspace{0.5cm}08 19 41.12   & $+$05 49 42.7   &   1.701   & 1.923  & 0.987  &  \hspace{0.5cm}S,N,F& \hspace{0.5cm}8 \\
J0842$+$2147       & \hspace{0.5cm}08 42 39.96   & $+$21 47 10.4   &   1.182   & 2.314  & 1.156  &  \hspace{0.5cm}S,N,F& \hspace{0.5cm}8 \\
J0902$+$5707       & \hspace{0.5cm}09 02 07.20   & $+$57 07 37.9   &   1.595   & 1.678  & 0.862  &  \hspace{0.5cm}S,N,F& \hspace{0.5cm}9, 8 \\
{\bf J0918$+$2325} & \hspace{0.5cm}09 18 58.15   & $+$23 25 55.4   &   0.688   & 2.079  & 0.885  &  \hspace{0.5cm}S,N,F& \hspace{0.5cm}10\\
{\bf J0925$+$4004} & \hspace{0.5cm}09 25 54.72   & $+$40 04 14.2   &   0.471   & 4.379  & 1.546  &  \hspace{0.5cm}S,N,F& \hspace{0.5cm}10\\
{\bf J0937$+$2937} & \hspace{0.5cm}09 37 04.04   & $+$29 37 04.8   &   0.451   & 2.640  & 0.909  &  \hspace{0.5cm}S,N,F& \hspace{0.5cm}6, 10 \\
{\bf J0944$+$2331} & \hspace{0.5cm}09 44 18.80   & $+$23 31 18.5   &   0.987   & 1.870  & 0.899  &  \hspace{0.5cm}S,N,F& \hspace{0.5cm}10\\
{\bf J0959$+$1216} & \hspace{0.5cm}09 59 34.49   & $+$12 16 31.6   &   1.089   & 1.964  & 0.966  &  \hspace{0.5cm}S,N,F& \hspace{0.5cm}11\\
                   &                             &                 &           &        &        &                     &                 \\
{\bf J1012$+$4229} & \hspace{0.5cm}10 12 44.29   & $+$42 29 57.0   &   0.364   & 3.088  & 0.933  &  \hspace{0.5cm}S,N,F& \hspace{0.5cm}9 \\
{\bf J1020$+$0447} & \hspace{0.5cm}10 20 26.87   & $+$04 47 52.0   &   1.131   & 1.478  & 0.733  &  \hspace{0.5cm}S,N,F& \hspace{0.5cm}11\\ 
{\bf J1020$+$3958} & \hspace{0.5cm}10 20 41.15   & $+$39 58 11.2   &   0.830   & 2.663  & 1.217  &  \hspace{0.5cm}W,N,F& \hspace{0.5cm}9 \\
J1027$-$2312       & \hspace{0.5cm}10 27 54.91   & $-$23 12 02.0   &   0.309   & 2.860  & 0.774  &  \hspace{0.5cm}N    & \hspace{0.5cm}3 \\
J1030$+$5310       & \hspace{0.5cm}10 30 50.91   & $+$53 10 28.6   &   1.197   & 1.698  & 0.749  &  \hspace{0.5cm}S,N,F& \hspace{0.5cm}8 \\
{\bf J1054$+$4152} & \hspace{0.5cm}10 54 03.27   & $+$41 52 57.6   &   1.090   & 4.702  & 2.314  &  \hspace{0.5cm}S,N,F& \hspace{0.5cm}10\\
{\bf J1056$+$4100} & \hspace{0.5cm}10 56 36.26   & $+$41 00 41.3   &   1.785   & 1.543  & 0.791  &  \hspace{0.5cm}S,N,F& \hspace{0.5cm}11\\
J1130$-$1320       & \hspace{0.5cm}11 30 19.90   & $-$13 20 50.0   &   0.634   & 4.812  & 1.977  &  \hspace{0.5cm}N    & \hspace{0.5cm}12\\
{\bf J1145$-$0033} & \hspace{0.5cm}11 45 53.67   & $-$00 33 04.6   &   2.054   & 2.642  & 1.340  &  \hspace{0.5cm}S,N,F& \hspace{0.5cm}13\\
J1148$-$0403       & \hspace{0.5cm}11 48 55.89   & $-$04 04 09.6   &   0.341   & 3.265  & 0.945  &  \hspace{0.5cm}N,F  & \hspace{0.5cm}14\\
                   &                             &                 &           &        &        &                     &                 \\
{\bf J1151$+$3355} & \hspace{0.5cm}11 51 39.68   & $+$33 55 41.8   &   0.851   & 2.083  & 0.959  &  \hspace{0.5cm}S,N,F& \hspace{0.5cm}10\\
{\bf J1229$+$3555} & \hspace{0.5cm}12 29 25.56   & $+$35 55 32.5   &   0.828   & 1.672  & 0.761  &  \hspace{0.5cm}S,N,F& \hspace{0.5cm}15\\
{\bf J1304$+$2454} & \hspace{0.5cm}13 04 51.42   & $+$24 54 45.9   &   0.605   & 2.431  & 0.977  &  \hspace{0.5cm}W,N,F& \hspace{0.5cm}10\\
{\bf J1321$+$3741} & \hspace{0.5cm}13 21 06.42   & $+$37 41 54.0   &   1.135   & 1.531  & 0.759  &  \hspace{0.5cm}S,N,F& \hspace{0.5cm}10\\
{\bf J1340$+$4232} & \hspace{0.5cm}13 40 34.70   & $+$42 32 32.2   &   1.343   & 2.309  & 1.173  &  \hspace{0.5cm}S,N,F& \hspace{0.5cm}10\\
J1353$+$2631       & \hspace{0.5cm}13 53 35.92   & $+$26 31 47.5   &   0.310   & 2.803  & 0.761  &  \hspace{0.5cm}W,N,F& \hspace{0.5cm}10, 16\\
{\bf J1408$+$3054} & \hspace{0.5cm}14 08 06.21   & $+$30 54 48.5   &   0.837   & 3.618  & 1.658  &  \hspace{0.5cm}S,N,F& \hspace{0.5cm}10\\
{\bf J1410$+$2955} & \hspace{0.5cm}14 10 36.80   & $+$29 55 50.9   &   0.570   & 2.483  & 0.970  &  \hspace{0.5cm}W,N,F& \hspace{0.5cm}6 \\
J1427$+$2632       & \hspace{0.5cm}14 27 35.61   & $+$26 32 14.5   &   0.363   & 3.822  & 1.158  &  \hspace{0.5cm}S,N,F& \hspace{0.5cm}16\\
J1432$+$1548       & \hspace{0.5cm}14 32 15.54   & $+$15 48 22.4   &   1.005   & 2.824  & 1.364  &  \hspace{0.5cm}S,N,F& \hspace{0.5cm}14\\
                   &                             &                 &           &        &        &                     &                 \\
J1504$+$6856       & \hspace{0.5cm}15 04 12.77   & $+$68 56 12.8   &   0.318   & 3.140  & 0.867  &  \hspace{0.5cm}N    & \hspace{0.5cm}5 \\
J1723$+$3417       & \hspace{0.5cm}17 23 20.80   & $+$34 17 58.0   &   0.206   & 3.787  & 0.760  &  \hspace{0.5cm}W,N,F& \hspace{0.5cm}17\\
J2042$+$7508       & \hspace{0.5cm}20 42 37.30   & $+$75 08 02.5   &   0.104   &10.052  & 1.138  &  \hspace{0.5cm}N    & \hspace{0.5cm}18\\
J2234$-$0224       & \hspace{0.5cm}22 34 58.76   & $-$02 24 18.9   &   0.550   & 3.236  & 1.241  &  \hspace{0.5cm}N,F  & \hspace{0.5cm}1 \\
{\bf J2344$-$0032} & \hspace{0.5cm}23 44 40.04   & $-$00 32 31.7   &   0.503   & 2.658  & 0.973  &  \hspace{0.5cm}W,N,F& \hspace{0.5cm}1 \\
\hline
\end{tabular}\\
References:(1) \cite{b4}; (2) \cite{b38}; (3) \cite{b25}; (4) \cite{b54}; (5) \cite{b34}; (6) \cite{b20}; (7) \cite{b55}; (8) \cite{b32}; (9) \cite{b65}; (10) \cite{b69}; (11) \cite{b31}; (12) \cite{b5}; (13) \cite{b33}; (14) \cite{b22}; (15) \cite{b57}; (16) \cite{b47}; (17) \cite{b26}; (18) \cite{b53}.
\end{minipage}
\end{table*}

\begin{table*}
 \centering
 \begin{minipage}{125mm}
  \caption{List of smaller ($<$ 0.72 Mpc) radio quasars.}
  \begin{tabular}{@{}lccccccl@{}}
  \hline
IAU          & \hspace{0.5cm}$\alpha$(J2000.0)       & $\delta$(J2000.0)&       z          &  d     & D      &\hspace{0.5cm}Avail.  &   Ref.\\
name         & \hspace{0.5cm}(h m s)               & ($^{o}$ \arcmin\, \arcsec)  &                   & arcmin & Mpc    & \hspace{0.5cm}Data       &  \\
(1)          & \hspace{0.5cm}(2)                   & (3)            &        (4)       &(5)     & (6)    & \hspace{0.5cm}(7)    &  (8) \\
\hline
J0022$-$0145  &       \hspace{0.5cm}00 22 44.29    &$-$01 45 51.1   &        0.691     & 1.432 & 0.610  &\hspace{0.5cm}N,F   &  1\\
J0034$+$0118  &       \hspace{0.5cm}00 34 19.18    &$+$01 18 35.8   &        0.841     & 1.364 & 0.664  &\hspace{0.5cm}W,N,F &  1\\
J0051$-$0902  &       \hspace{0.5cm}00 51 15.12    &$-$09 02 08.5   &        1.265     & 1.379 & 0.696  &\hspace{0.5cm}S,N,F &  1\\
J0130$-$0135  &       \hspace{0.5cm}01 30 43.00    &$-$01 35 08.2   &        1.160     & 1.306 & 0.650  &\hspace{0.5cm}W,N,F &  1 \\
J0245$+$0108  &       \hspace{0.5cm}02 45 34.07    &$+$01 08 14.2   &        1.537     & 0.883 & 0.453  &\hspace{0.5cm}S,N,F &  3\\
J0745$+$3142  &      \hspace{0.5cm} 07 45 41.66    &$+$31 42 56.5   &        0.461     & 1.795 & 0.626  &\hspace{0.5cm}S,N,F &  3\\
J0811$+$2845  &       \hspace{0.5cm}08 11 36.90    &$+$28 45 03.6   &        1.890     & 0.507 & 0.259  &\hspace{0.5cm}S,N,F &  3\\
J0814$+$3237  &       \hspace{0.5cm}08 14 09.23    &$+$32 37 31.7   &        0.844     & 0.239 & 0.187  &\hspace{0.5cm}S,N,F &  3\\
J0817$+$2237  &       \hspace{0.5cm}08 17 35.07    &$+$22 37 18.0   &        0.982     & 0.395 & 0.190  &\hspace{0.5cm}S,N,F &  3\\
J0828$+$3935  &       \hspace{0.5cm}08 28 06.85    &$+$39 35 40.3   &        0.761     & 1.077 & 0.477  &\hspace{0.5cm}S,N,F &  3\\
              &                                    &                &                  &       &        &                    &   \\
J0839$+$1921  &       \hspace{0.5cm}08 39 06.95    &$+$19 21 48.9   &        1.691     & 0.523 & 0.269  &\hspace{0.5cm}S,N,F &  3\\ 
J0904$+$2819  &       \hspace{0.5cm}09 04 29.63    &$+$28 19 32.8   &        1.121     & 0.379 & 0.188  &\hspace{0.5cm}S,N,F &  3\\  
J0906$+$0832  &       \hspace{0.5cm}09 06 49.81    &$+$08 32 58.8   &        1.617     & 1.307 & 0.671  &\hspace{0.5cm}S,N,F &  4\\
J0924$+$3547  &       \hspace{0.5cm}09 24 25.03    &$+$35 47 12.8   &        1.342     & 1.345 & 0.683  &\hspace{0.5cm}S,N,F &  5\\
J0925$+$1444  &       \hspace{0.5cm}09 25 07.26    &$+$14 44 25.9   &        0.896     & 0.665 & 0.311  &\hspace{0.5cm}S,N,F &  3\\
J0935$+$0204  &       \hspace{0.5cm}09 35 18.51    &$+$02 04 19.0   &        0.649     & 1.200 & 0.498  &\hspace{0.5cm}S,N,F &  3\\
J0941$+$3853  &       \hspace{0.5cm}09 41 04.17    &$+$38 53 49.1   &        0.616     & 0.853 & 0.346  &\hspace{0.5cm}S,N,F &  3\\
J0952$+$2352  &       \hspace{0.5cm}09 52 06.36    &$+$23 52 43.2   &        0.970     & 1.466 & 0.702  &\hspace{0.5cm}S,N,F &  2\\
J1000$+$0005  &       \hspace{0.5cm}10 00 17.65    &$+$00 05 23.9   &        0.905     & 0.521 & 0.245  &\hspace{0.5cm}S,N,F &  3\\ 
J1004$+$2225  &       \hspace{0.5cm}10 04 45.75    &$+$22 25 19.4   &        0.982     & 1.097 & 0.526  &\hspace{0.5cm}S,N,F &  3\\
              &                                    &                &                  &       &        &                    &   \\
J1005$+$5019  &       \hspace{0.5cm}10 05 07.10    &$+$50 19 31.5   &        2.023     & 1.300 & 0.660  &\hspace{0.5cm}S,N,F &  2\\
J1006$+$3236  &       \hspace{0.5cm}10 06 07.58    &$+$32 36 27.9   &        1.026     & 0.246 & 0.119  &\hspace{0.5cm}S,N,F &  3\\
J1009$+$0529  &       \hspace{0.5cm}10 09 43.56    &$+$05 29 53.9   &        0.942     & 1.377 & 0.654  &\hspace{0.5cm}S,N,F &  2\\
J1010$+$4132  &       \hspace{0.5cm}10 10 27.50    &$+$41 32 39.0   &        0.612     & 0.525 & 0.212  &\hspace{0.5cm}S,N,F &  3\\
J1023$+$6357  &       \hspace{0.5cm}10 23 14.61    &$+$63 57 09.3   &        1.194     & 1.294 & 0.648  &\hspace{0.5cm}S,N,F &  6\\
J1100$+$1046  &       \hspace{0.5cm}11 00 47.81    &$+$10 46 13.6   &        0.422     & 0.549 & 0.182  &\hspace{0.5cm}S,N,F &  3\\
J1100$+$2314  &       \hspace{0.5cm}11 00 01.14    &$+$23 14 13.1   &        0.559     & 1.577 & 0.610  &\hspace{0.5cm}S,N,F &  5\\
J1107$+$0547  &       \hspace{0.5cm}11 07 09.51    &$+$05 47 44.7   &        1.799     & 1.324 & 0.678  &\hspace{0.5cm}S,N,F &  2\\
J1107$+$1628  &       \hspace{0.5cm}11 07 15.04    &$+$16 28 02.2   &        0.632     & 0.652 & 0.267  &\hspace{0.5cm}S,N,F &  3\\
J1110$+$0321  &       \hspace{0.5cm}11 10 23.84    &$+$03 21 36.4   &        0.966     & 1.055 & 0.504  &\hspace{0.5cm}S,N,F &  3\\
              &                                    &                &                  &       &        &                    &   \\
J1118$+$3828  &       \hspace{0.5cm}11 18 58.53    &$+$38 28 53.5   &        0.747     & 1.407 & 0.619  &\hspace{0.5cm}S,N,F &  5\\
J1119$+$3858  &       \hspace{0.5cm}11 19 03.20    &$+$38 58 53.6   &        0.734     & 1.419 & 0.620  &\hspace{0.5cm}S,N,F &  5\\
J1158$+$6254  &       \hspace{0.5cm}11 58 39.76    &$+$62 54 27.1   &        0.592     & 0.968 & 0.385  &\hspace{0.5cm}S,N,F &  3\\ 
J1217$+$1019  &       \hspace{0.5cm}12 17 01.28    &$+$10 19 52.0   &        1.883     & 0.466 & 0.238  &\hspace{0.5cm}S,N,F &  3\\ 
J1223$+$3707  &       \hspace{0.5cm}12 23 11.23    &$+$37 07 01.8   &        0.491     & 0.597 & 0.216  &\hspace{0.5cm}S,N,F &  3\\
J1236$+$1034  &       \hspace{0.5cm}12 36 04.52    &$+$10 34 49.2   &        0.667     & 1.694 & 0.711  &\hspace{0.5cm}S,N,F &  3\\ 
J1256$+$1008  &       \hspace{0.5cm}12 56 07.66    &$+$10 08 53.5   &        0.824     & 0.382 & 0.174  &\hspace{0.5cm}S,N,F &  3\\
J1319$+$5148  &       \hspace{0.5cm}13 19 46.25    &$+$51 48 05.5   &        1.061     & 0.466 & 0.228  &\hspace{0.5cm}S,N,F &  3\\
J1334$+$5501  &       \hspace{0.5cm}13 34 11.71    &$+$55 01 24.8   &        1.245     & 1.274 & 0.641  &\hspace{0.5cm}S,N,F &  3\\
J1358$+$5752  &       \hspace{0.5cm}13 58 17.60    &$+$57 52 04.5   &        1.373     & 0.733 & 0.373  &\hspace{0.5cm}S,N,F &  3\\
              &                                    &                &                  &       &        &                    &   \\
J1425$+$2404  &       \hspace{0.5cm}14 25 50.65    &$+$24 04 02.8   &        0.653     & 0.339 & 0.141  &\hspace{0.5cm}S,N,F &  3\\
J1433$+$3209  &       \hspace{0.5cm}14 33 34.26    &$+$32 09 09.5   &        0.935     & 0.630 & 0.299  &\hspace{0.5cm}S,N,F &  3\\ 
J1513$+$1011  &       \hspace{0.5cm}15 13 29.30    &$+$10 11 05.4   &        1.546     & 0.586 & 0.301  &\hspace{0.5cm}S,N,F &  3\\
J1550$+$3652  &       \hspace{0.5cm}15 50 02.01    &$+$36 52 16.8   &        2.061     & 1.334 & 0.676  &\hspace{0.5cm}S,N,F &  4\\
J1557$+$0253  &       \hspace{0.5cm}15 57 52.83    &$+$02 53 28.9   &        1.988     & 1.121 & 0.571  &\hspace{0.5cm}S,N,F &  2\\
J1557$+$3304  &       \hspace{0.5cm}15 57 29.94    &$+$33 04 47.0   &        0.953     & 0.562 & 0.268  &\hspace{0.5cm}S,N,F &  3\\ 
J1622$+$3531  &       \hspace{0.5cm}16 22 29.90    &$+$35 31 25.1   &        1.475     & 0.365 & 0.187  &\hspace{0.5cm}S,N,F &  3\\
J1623$+$3419  &       \hspace{0.5cm}16 23 36.45    &$+$34 19 46.3   &        1.981     & 0.984 & 0.501  &\hspace{0.5cm}S,N,F &  2\\
J2335$-$0927  &       \hspace{0.5cm}23 35 34.68    &$-$09 27 39.2   &        1.814     & 1.305 & 0.668  &\hspace{0.5cm}S,N,F &  1\\
\hline 
\end{tabular}
References: (1) \cite{b4}; (2) \cite{b65}; (3) \cite{b47}; (4) \cite{b32}; (5) \cite{b69}; (6) \cite{b31}.
\end{minipage}
\end{table*}

\section{Sample Biases}

Due to the method used to complete our sample, the results may in some cases be influenced by selection effects, for example related to the sensitivity of the radio surveys used for selecting extended sources. The sample of giant radio quasars was compiled in  three stages and each of them may be affected by bias. First, compact radio objects were selected, then the optical counterparts were checked for  spectra typical of quasars. The selection criteria for these steps were described in detail in the papers referenced in Sect. 2 and we will not focus on them here. In the third stage of selection, we inspected the radio maps of several hundred candidates, looking for targets which have extended radio lobes in addition to radio cores. The NVSS and FIRST surveys have a completeness of 96 and 89 per cent and a reliability of 99 and 94 per cent to the $5\sigma$ limits of 2.3 and 1.0 mJy respectively (\citealt{b1131}). Since the resolution effect causes FIRST to become more incomplete for extended objects, we supplemented our search with the NVSS maps which have larger restoring beam size and hence larger surface brightness sensitivity. However, because of the limited baselines NVSS is insensitive to very extended coherent structures (larger than 15\arcmin). Fortunately, we do not expect the existence of objects with such large angular size, at least at high redshifts. In addition, extended and aged radio sources could have weak double lobes not connected with a visible bridge of high frequency radio emission. Therefore, it may be hard to recognise such a source as just one homogeneous object, especially at high redshift where the inverse Compton losses against the cosmic microwave background are large. Detecting a steep-spectrum and low surface-brightness radio bridge connecting the radio core with hot spots for distant objects is therefore challenging and this may have caused us to overlook some objects.

It is worth noting that most of the recent works on quasars based on optical and radio data first select the candidates from optical catalogues of quasars and then correlate their coordinates with catalogues of radio sources. The authors usually concentrate on point-like radio sources, not extended objects (there are some exceptions, however, for example \citealt{b65}). \cite{b1123} considered extended radio structures, but analysed only those objects whose lobe separation  was smaller than 1\arcmin. The authors stressed that the extended radio quasars represented a very small fraction of the SDSS quasars. They also wrote that quasars for which the radio structure diameter is greater then 1\arcmin are even rarer. Thus one has to realize that objects of the class studied here are extremely rare.

As we pointed out in Sect.~2, the lobe-dominated radio quasars lie almost in the plane of the sky. Therefore, their measured radio luminosity is weakly influenced by relativistic beaming. In addition, it is easy to determine the proper physical size and volume occupied by the radio plasma for sources oriented in this manner. On the other hand, one should keep in mind that, besides lobe-dominated giant radio quasars as focused on here, there exist giant radio quasars located at a small angle to the line of sight which we have completely ignored because of the inability to determine their physical size.

Given all the drawbacks described above, we have nonetheless shown that giant radio quasars do not comprise just a few objects as previously thought, but constitute a larger group. In addition to the sample of newly identified giants, we also added the set of previously known giant quasars to increase the number of objects tested. Summing up, the sample we presented here is limited by the described selection criteria and is not fully homogeneous. Therefore, applying the conclusions obtained here to the whole population of radio-loud quasars should be done with caution.

\section[]{Data analysis}
\subsection[]{Radio data}
Using the Astronomical Image Processing System\footnote {http://www.aips.nrao.edu/}  package for radio data reduction and analysis and maps from the NVSS and FIRST surveys, we measured the basic parameters of the selected radio quasars, which were further used to calculate their characteristics -- defined in the following way:

\begin{enumerate}
\item The arm-length-ratio, $Q$, which is the ratio of distances ($d_1$ and $d_2$) between the core and the hot spots (peaks of radio emission), normalized in such a way that always $Q>1$ (for details see Fig.~1).

\item The bending angle, $B$, which is the complement of the angle between the lines connecting the lobes with the core.

\item The lobes' flux-density ratio, $F=S_1/S_2$, where $S_1$ is the flux density of the lobe further from the core and $S_2$ is the flux density of the lobe closer to the core.

\item  The source total luminosity at 1.4~GHz, $P\rm_{tot}$, which is calculated following the formula given by \cite{b12}:

 \begin{eqnarray} 
    \lefteqn{log P_{tot}(WHz^{-1})=log S_{tot}(mJy)-(1+\alpha) \cdot log(1+z)}
    \nonumber\\ 
    & & {}+2log(D_L(Mpc))+17.08 
 \label{eq1}
 \end{eqnarray}

\noindent
where $\alpha$ is the spectral index (the convention we use here is $S_{\nu}\sim\nu^{\alpha}$) and $D_L$ is a luminosity distance. The total flux density, $S_{\rm tot}$, of individual sources is measured from NVSS maps and the average spectral index, in accordance with \cite{b68}, is taken for all sources as $\alpha=-0.6$. The core luminosity at 1.4~GHz, $P\rm_{core}$, is calculated in a similar manner, but instead of $S_{\rm tot}$ in equation~(\ref{eq1}) we substitute the core flux density, $S_{\rm core}$, which is measured from FIRST maps and the average spectral index value, according to \cite{b72}, is adopted as $\alpha = -0.3$.
  
\item The inclination angle, $i$, which is the angle between the jet axis and the line of sight (i.e. $i = 90$\degr means that the object lies in the sky plane). The inclination angle was calculated, assuming that the Doppler boosting is the main factor underlying the asymmetries of a source, in the following way:

 \begin{equation}
   i=[acos(\frac{1}{\beta_j} \cdot \frac{(s-1)}{(s+1)})] 
 \label{eq2}
 \end{equation}

\noindent
where $s=(S_j/S_{cj})^{1/2-\alpha}$, $S_j$ is the peak flux-density of the lobe closer to the core. $S_{cj}$ is the peak flux-density of the lobe further from the core and $\beta_j$ is the jet velocity. For all our objects, according to \cite{b68} and \cite{b1129}, we assume $\beta_j=0.6$c.
\end{enumerate} 

\noindent
The resulting values of the above parameters for our sources are listed in Table 3. For two objects, i.e. J0439$-$2422 and J1100$+$2314, we were not able to measure all the parameters, as for the source J0439$-$2422 the FIRST map was not available, and J1100$+$2314 has a too asymmetric radio structure. 

\begin{figure}
\centering
 \includegraphics[width=0.75\linewidth, angle=0]{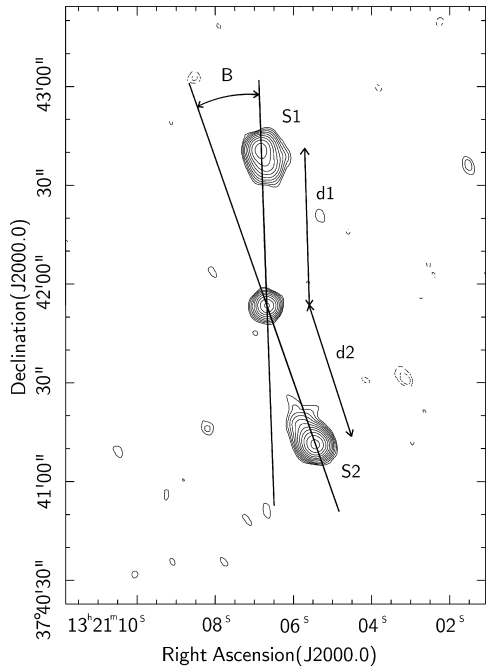}
 \caption{An example of a GRQ, J1321$+$3741. Radio contours are taken from the FIRST survey. Definitions of some parameters used for analysis are provided here (i.e. $B$, $S_1$, $S_2$, $d_1$, $d_2$) and also described in the text.}
\end{figure}

\subsection[]{Optical data}
\subsubsection[]{Spectra reduction}
The quasar spectra were reduced through the standard procedures of the Image Reduction and Analysis Facility \footnote {http://iraf.noao.edu/} package including galactic extinction and redshift correction. Each spectrum was corrected for galactic extinction taking into account values of the colour excess $E(B-V)$ and the $B$-band extinction, $A_B$, taken from the NASA/IPAC Extragalactic Database. We calculated the extinction parameter $R=E(B-V)/A_B$ for each quasar in our samples. The extinction-corrected spectrum was then transformed to its rest frame using the redshift value given in the SDSS (or if the SDSS spectrum was unavailable, from other publications).  
    
\subsubsection[]{Continuum subtraction and line parameters measurement}
In order to obtain reliable measurements of emission lines, we need to subtract continuum emission, as optical and UV spectra of quasars are dominated by the power-law and Balmer continuum. Using the Image Reduction and Analysis Facility package, we subtracted the power-law continuum from our spectra. The continuum was fitted in several windows where we had not observed any emission lines (i.e. 1320--1350\AA, 1430--1460\AA, 1790--1830\AA, 3030--3090\AA, 3540--3600\AA\,  and 5600--5800\AA). Particularly in the UV band, we also observe significant iron emission, which is often blended with the MgII(2798\AA) line. The procedure of subtracting the iron emission was similar to that described by \cite{b11}. We used an Fe template in the UV band (1250--3090\AA) as developed by \cite{b64}, and in the optical band (3535--7530\AA) given by \cite{b60}. First, we broadened the iron template by convolving it with Gaussian functions of various widths and multiplying by a scalar factor. Next, we chose the best fit of this modified template to each particular spectrum, and then subtracted it. After the subtraction of Fe line emission, we added the previously determined power-law continuum fit and refitted it once again (in a similar manner as suggested by \citealt {b64}). An example of a ``cleaned-up'' spectrum is presented in Fig.~2.\\
For the purpose of our analysis we needed to measure the parameters of broad emission lines like CIV(1549\AA), MgII(2798\AA) and H$_\beta$(4861\AA). In some cases, performing this measurement was difficult due to asymmetries in the line profiles (particularly of highly ionized lines such as CIV), where it was hard to fit a Gaussian function. In order to  overcome the problem, we used the method described in \cite{b50}. In Tables 4 and 5 (cols.~2--4) we provide the respective widths of broad emission lines for GRQs and smaller quasars, respectively. We were unable to measure the MgII emission line parameters in the spectrum of the GRQ J1408+3054, as it showed strong broad-absorption features which considerably affected the emission line profile.  

\subsubsection[]{Black hole mass determination}
The issue of determination of BH mass in AGNs has recently been often studied. The knowledge of the BH mass is of great importance in determining a number of physical parameters of AGNs and their evolution. In the first place, all of the commonly known techniques based on kinematic or dynamical studies  (e.g. \citealt{b52}) are only useful for inactive galaxies. Therefore, they cannot be applied directly for AGNs, which are very luminous and distant. The most promising method for AGNs is reverberation mapping of the broad emission lines from the broad-line region (\citealt{b49}). This method works particularly well for type I AGNs (e.g. \citealt{b66}), where the broad line region is not obscured by a dusty and gaseous torus. Assuming that the gas in the broad-line region is virialized in the gravitational field of a BH, we can calculate its mass as:

\begin{equation}
     M_{\rm BH}=\frac{R_{\rm BLR} V^2_{\rm BLR}}{\rm G}
\label{eq3}
\end{equation}

\noindent
where G is the gravitational constant, $R_{\rm BLR}$ is the distance from broad-line region clouds to the central BH, $V_{\rm BLR}$ is the broad-line region virial velocity, which can be estimated from the FWHM (Full Width at Half Maximum) of a respective emission line  as:

\begin{equation}
     V_{\rm BLR}=f \cdot FWHM
\label{eq4}
\end{equation}

\noindent
where $f$ is the scaling factor, which depends on structure, kinematics, and orientation of the broad line region (for randomly distributed broad line region clouds $f=\sqrt{3}/2$). Basing on this method, \cite{b28, b29} obtained an empirical relation between the broad line region size of an AGN and its optical continuum luminosity ($\lambda L_{\lambda}$) at 5100$\rm\AA$ (and later also at 1450$\rm\AA$, 1350$\rm\AA$ and in the 2--10~keV range):

\begin{equation}
     R_{\rm BLR} \sim \lambda L_{\lambda}(5100\rm\AA)^{0.70 \pm 0.03}
\label{aga5}
\end{equation}

\noindent
This relation makes it possible to use an approximation to the reverberation mapping method, called the mass-scaling relation, which allows to determine BH mass using measurements of the FWHM of broad emission lines (e.g. CIV, MgII, H$\beta$) and the monochromatic continuum luminosity ($\lambda L_{\lambda}$) of a single-epoch spectrum only. In order to determine BH mass of our objects basing on the FWHM measurements of different emission lines, we applied the following equations:

\begin{eqnarray}
    \lefteqn{M_{\rm BH}(CIV1549\rm\AA)= 4.57 \cdot 10^6 (\frac{\lambda
L_{\lambda}(1350\rm\AA)}{10^{44}erg s^{-1}})^{0.53\pm0.06} \cdot}
    \nonumber\\
    & & {}(\frac{FWHM(CIV1549\rm\AA)}{1000 km s^{-1}})^2 M_{\odot}
\label{eq6}
\end{eqnarray}

\begin{eqnarray}
    \lefteqn{M_{\rm BH}(MgII2798\rm\AA)=7.24 \cdot 10^6(\frac{\lambda
L_{\lambda}(3000\rm\AA)}{10^{44}erg s^{-1}})^{0.5} \cdot}
    \nonumber\\
    & & {}(\frac{FWHM(MgII2798\rm\AA)}{1000 km s^{-1}})^2 M_{\odot}
\label{eq7}
\end{eqnarray}

\begin{eqnarray}
  \lefteqn{M_{\rm BH}(H\beta4861\rm\AA)= 8.13 \cdot 10^6 (\frac{\lambda
L_{\lambda}(5100\rm\AA)}{10^{44}erg s^{-1}})^{0.50\pm0.06} \cdot}
    \nonumber\\
    & & {}(\frac{FWHM(H\beta4861\rm\AA)}{1000 km s^{-1}})^2 M_{\odot}
\label{eq8}
\end{eqnarray}

\noindent
Equations (\ref{eq6}) and (\ref{eq8}) were taken from \cite{b63}, while equation (\ref{eq7}) from \cite{b62}. The monochromatic continuum luminosities $\lambda L_{\lambda}$ can be computed as follows:

\begin{equation}
    \lambda L_{\lambda} = 4 \pi D_{\rm Hubble}^2 \lambda f_{\lambda}
\label{eq9}
\end{equation}

\noindent
where $D_{\rm Hubble}$ is the comoving radial distance and $f_{\lambda}$ is the flux in the rest frame at wavelength $\lambda$ equal to 3000\AA, 5100\AA, or 1350\AA. The resulting rest frame fluxes, monochromatic continuum luminosities and BH masses for GRQs and smaller quasars are given in Table 4 and 5 (cols. 5--7, 8--10 and 11--13) respectively.\\

\begin{figure}
 \includegraphics[width=0.99\linewidth]{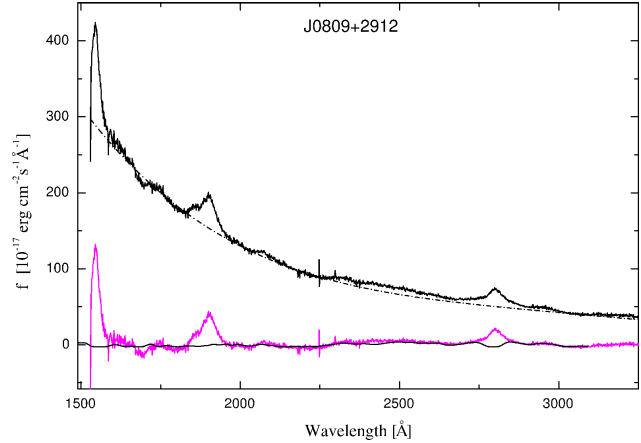}
 \caption{Spectrum of the giant radio quasar J0809+2912 and the best fit to the iron emission. The top spectrum is the observed spectrum in the rest frame overlaid with a power-law continuum, while the bottom one is the continuum-subtracted spectrum overlaid with the best fit to the iron emission.}
\end{figure}

\section[]{Results}
\subsection[]{Radio properties}
In our analysis we checked some general relations between radio parameters for our sample sources, similar to those shown for the sample of GRSs (mostly galaxies) described by \cite{b25}. On the optical- versus radio-luminosity plane our objects trace the regime of radio loudness (ratio of radio-to-optical luminosity) between 50 and 1000 and overlap with the FIRST-2dF sample of quasars of \cite{b1122}.

In Fig.~3 we present the dependence between 1.4~GHz total luminosity and redshift for our quasars. It is important to note that our comparison sample of smaller radio quasars (sources marked as open circles in Fig.~3 and subsequent figures) contains only objects in the redshift range of 0.4$\lesssim$z$\lesssim$2 due to our selection criteria, i.e. the presence of the MgII(2798\AA) emission line in the spectra (for details see Sect.~2). Such a cut-off in the redshift range of the quasars from the comparison sample should not, however, affect our main results, since the majority of GRQs have redshifts in a similar range. Therefore, the non-existence of smaller radio quasars in the upper-left part of Fig.~3 is artificial, whereas the absence of GRQs in the lower-right corner of this figure is the result of sensitivity limit of the radio surveys which we used for source recognition and measurements of source's radio properties. It is known that in flux-limited samples we should expect a correlation between radio luminosity and redshift, since for larger distances we are able to detect only those sources which are luminous enough, and faint sources at higher redshifts are beyond the detection limit. For our quasar sample a dependence between redshift and total radio luminosity can be seen, but the correlation is not as strong as for the sample of GRSs from \cite{b25}. The Spearman rank correlation coefficient for the GRQs is 0.49, whereas for the GRSs from the paper cited above it is 0.90. This shows that the selection effects for our quasar sample are not as strong as for other radio galaxies and GRS samples of \cite{b25}, though they may still have affected some of our results.

\begin{figure}
 \includegraphics[width=0.99\linewidth]{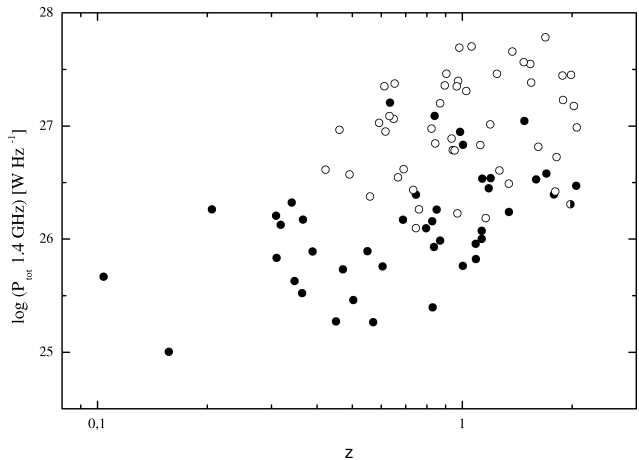}
 \caption{1.4 GHz total radio luminosity as a function of redshift. The GRQs are marked  with solid circles and quasars from the comparison sample are marked by open circles. J1623+3419, which is marked by a half-solid circle, has a projected linear size of 0.5 Mpc but, after correction for the inclination angle, its unprojected linear size is larger than the defining minimum size of GRSs.}
\end{figure} 

In Fig.~4 we present the luminosity, $P$, versus linear size, $D$, relation. The $P$--$D$ diagram is a helpful tool in investigating the evolution of radio sources and was frequently used to test evolutionary models (e.g. \citealt{b27, b9}). In order to draw this diagram we used the unprojected linear size of the sources, which was derived by taking into account the inclination angle, $i$, as $D^* = D/sin(i)$, where $D$ is the projected linear size (given in Tables 1 and 2 derived as the sum of $d_1$ and $d_2$ - for details see Fig.~1). The diagrams show that GRQs have, on average, lower core and total radio luminosities. The trend which we observe in our $P$--$D$ diagrams is consistent with the predictions of evolutionary models and can suggest that, under favourable conditions, the luminous smaller, and probably younger, radio quasars may evolve in time into the lower-luminosity aged GRQs. The non-existence of objects in the bottom-left part of Fig.~4 may be due to selection effects. Because of the surface-brightness limit we may overlook some extended objects with low total radio luminosities.
\begin{figure}
 \includegraphics[width=0.99\linewidth]{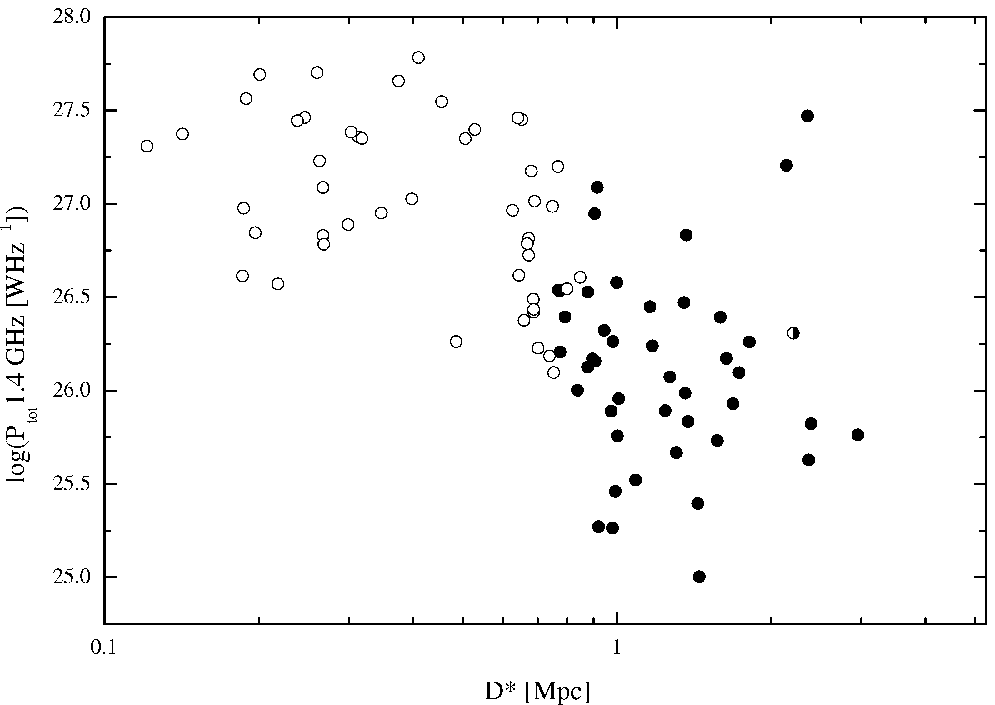}\\
 \includegraphics[width=0.99\linewidth]{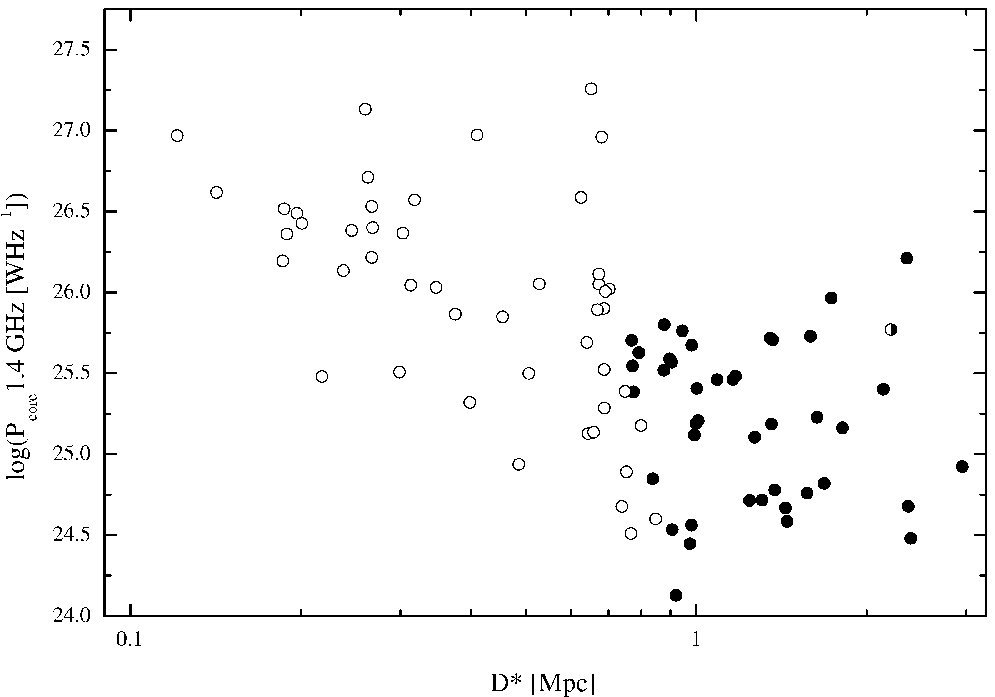}
 \caption{Luminosity--linear size diagrams. The top panel shows the 1.4~GHz total radio luminosity and the bottom one shows core luminosity. The observed trend is consistent with predictions of evolutionary models.}
\end{figure}

In Fig.~5 we present the relation between the total and core radio luminosity. There is a strong correlation between those two quantities for radio quasars. We obtained a correlation coefficient of 0.76 and the slope of the linear fit equal to $0.82 \pm 0.08$, steeper than the slope of  $0.59 \pm 0.05$ obtained by \cite{b25} for GRSs. The strong correlation between the core luminosity and the total luminosity in the population of giant-size radio galaxies was also mentioned by \cite{b37b}. On the one hand, this correlation can be attributed to the Doppler beaming of a parsec-scale jet and can reflect the different inclination angle of the nuclear jets, and thus inclination of the entire radio source's axis to the observer's line of sight. Relatively more luminous cores (in comparison to the total luminosity) should be observed in more strongly projected sources (i.e. quasars). Therefore, in GRQs one could expect to observe relatively stronger cores than in giant-sized radio galaxies. On the other hand, evolutionary effects (well visible in Fig.~4) can explain the clear difference in radio luminosity between GRQs and smaller quasars.

Some authors (e.g. \citealt{b29}) have suggested that giants should have more prominent cores, as stronger nuclear activity is necessary to produce the larger linear sizes of their radio structure. \cite{b25} attempted to verify this hypothesis for giant-sized radio galaxies by plotting a diagram of the core prominence, $f_c$, which is the ratio of core luminosity to the total luminosity of the radio source, but found no trend of this kind. For GRQs investigated in this paper we also plotted such a diagram (see Fig.~6) and came to a similar conclusion. We can reconcile this with the existence of the core luminosity -- total luminosity correlation visible in Fig.~5 as a result of smaller quasars having more luminous cores but also larger total luminosities than GRQs. The resulting mean values of $f_c$ are 0.20 and 0.18 for GRQs and smaller quasars respectively. In Fig.~7 we plot the core prominence against linear size of the extended radio structure. The distribution of the core prominence is similar for GRQs and smaller radio quasars, which allows the claim that the strength of the central engine of GRQs is similar to that of smaller radio quasars.
  
\begin{figure}
 \includegraphics[width=0.99\linewidth]{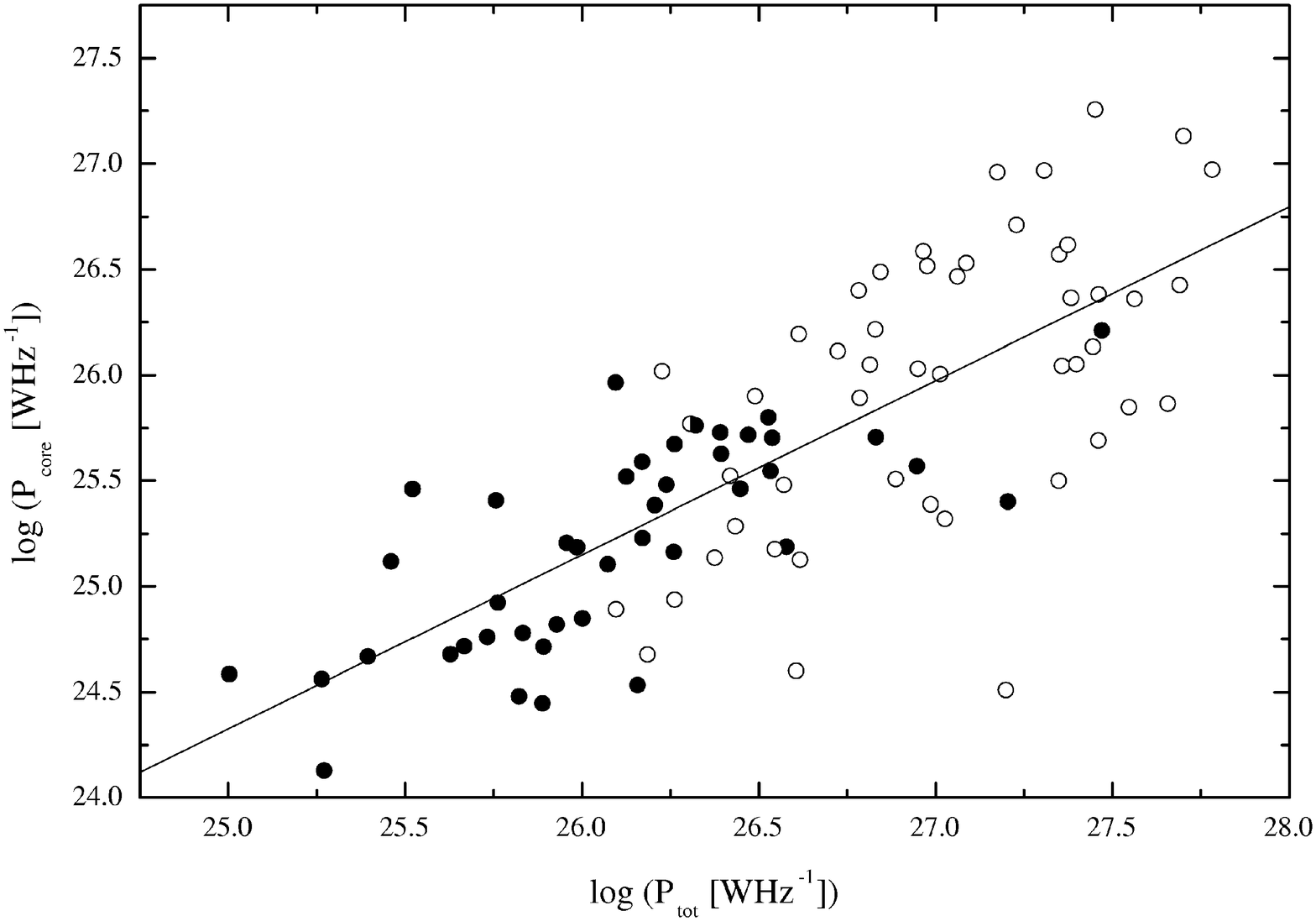}
 \caption{Core radio luminosity against the total radio luminosity for radio quasars. A strong correlation is visible. A linear fit to the data points is given by the line $logP_{\rm core}=(0.823\pm0.075)logP_{\rm tot}+(3.723\pm2.005)$.}
\end{figure}
\begin{figure}
 \includegraphics[width=0.99\linewidth]{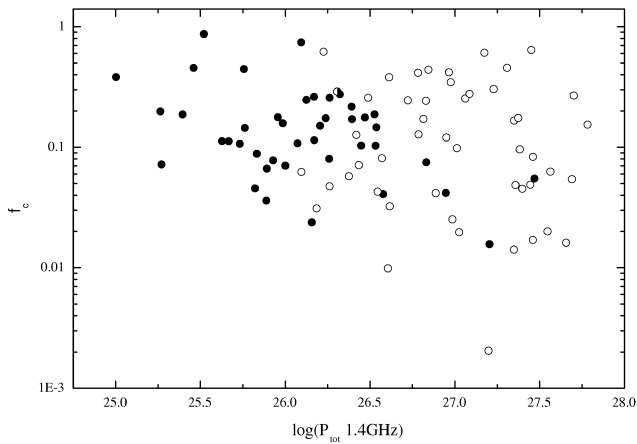}
 \caption{Core prominence against total radio luminosity.} 
\end{figure}
\begin{figure}
 \includegraphics[width=0.99\linewidth]{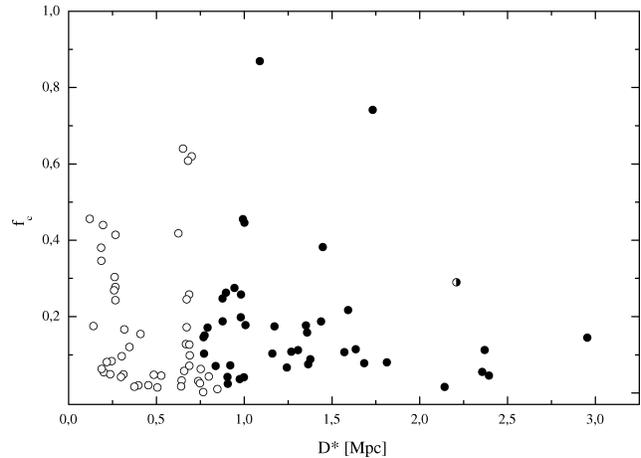}
 \caption{Core prominence against unprojected linear size. Also here, similar as in Fig~6, any correlation is visible.}
\end{figure}

We also investigated the asymmetries of radio structures in both our lobe-dominated radio quasar samples. It is well known that non-uniform environment (i.e. non-uniform density on both sides of the core) is one of the factors underlying radio structure asymmetries, which can be described by the arm-length ratio $Q$ (e.g. \citealt{bsch}). The distribution of this parameter for the GRQs and the quasars from the comparison sample is presented in Fig.~8. We found that the GRQs seem to be more symmetric than the smaller radio quasars (there were no GRQs with $Q > 2.4$ in our sample). However, the obtained mean values of the $Q$ parameter for GRQs and for the comparison sample are $1.41 \pm 0.33$ and $1.65 \pm 0.61$, respectively, and therefore indistinguishable within the error limits. This suggests that the IGM in which the giants evolve is not more symmetrical than that around the smaller sources. Our results for the large radio sources are comparable to those by \cite{b25}, which found that the mean value of the $Q$ parameter for the GRSs is 1.39, but for a comparison sample based on smaller 3CR sources they obtained a smaller $Q$ value equal to 1.19. We also compared the arm-length ratio value of GRQs and GRSs (see the bottom panel in Fig.~8). It can be seen clearly that the distributions for giant quasars and galaxies are similar.

The values of the bending angle $B$ and lobe flux-density ratio $F$ give similar result for both samples of quasars with mean values of  $B=7.40 \pm 5.89$, $F=1.45 \pm 1.16$ and $B=8.50 \pm 7.31$, $F=2.28 \pm 5.53$ for GRQs and comparison sample, respectively.

In summary, there is no significant difference in the environmental properties of the IGM within which giant- and smaller-sized radio quasars evolve.
 
Furthermore, we checked distribution of the inclination angle, $i$ (see Fig.~9). For our sample of radio quasars, we obtained that most objects have inclinations between 60$^{ \rm o}$ and 90$^{ \rm o}$. This result is inconsistent with the models of AGN unification scheme, where -- following \cite{b67} -- the inclination angle for quasars has a value between 0$^{ \rm o}$--45$^{ \rm o}$. In the objects with the angle larger than 45$^{ \rm o}$, the broad-line region should be partially or totally obscured by a dusty torus and the broad emission lines should not be as prominent as we observe in the spectra from our quasar sample. A plausible explanation of the observed distribution of inclinations is that there is no dusty torus in some AGNs (\citealt{b18}) or we are dealing with a clumpy, or receding torus (i.e. \citealt{b2222}), thus broad emission lines could have been observed even in quasars with large inclinations. The quasar with the largest asymmetry of its radio structure is J1623+3419 with $i=13^{ \rm o}$. Such a small value of the inclination angle can suggest that it should rather be classified as a BL~Lac object. Further observations are needed to confirm if its observed radio structure is actually related to a unique radio source.

\begin{figure}
 \includegraphics[width=0.99\linewidth]{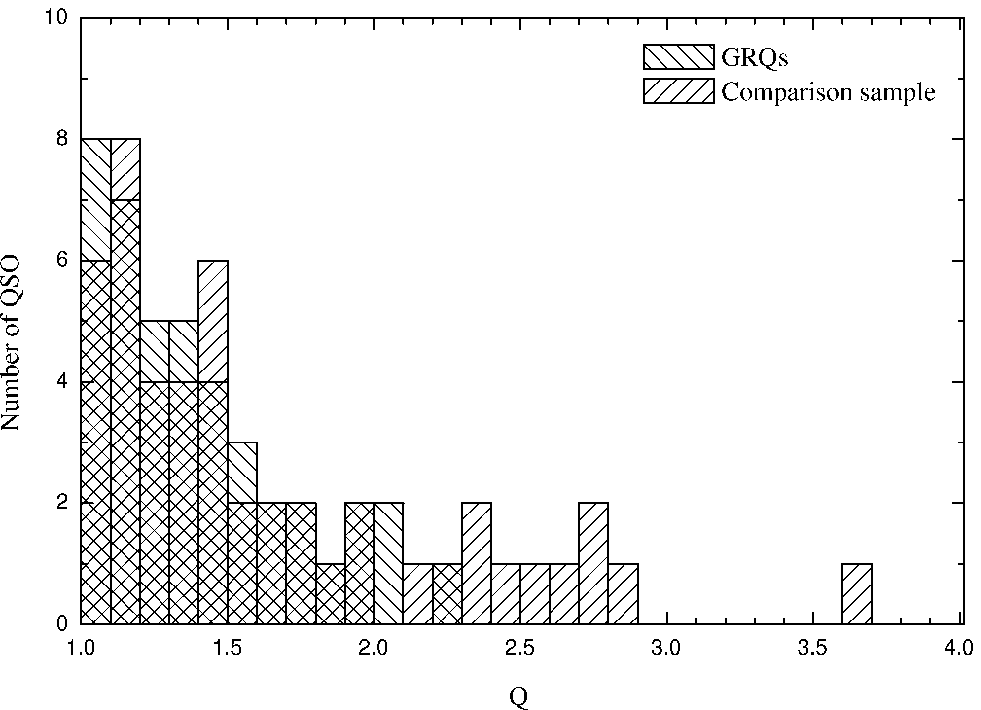}
 \includegraphics[width=0.99\linewidth]{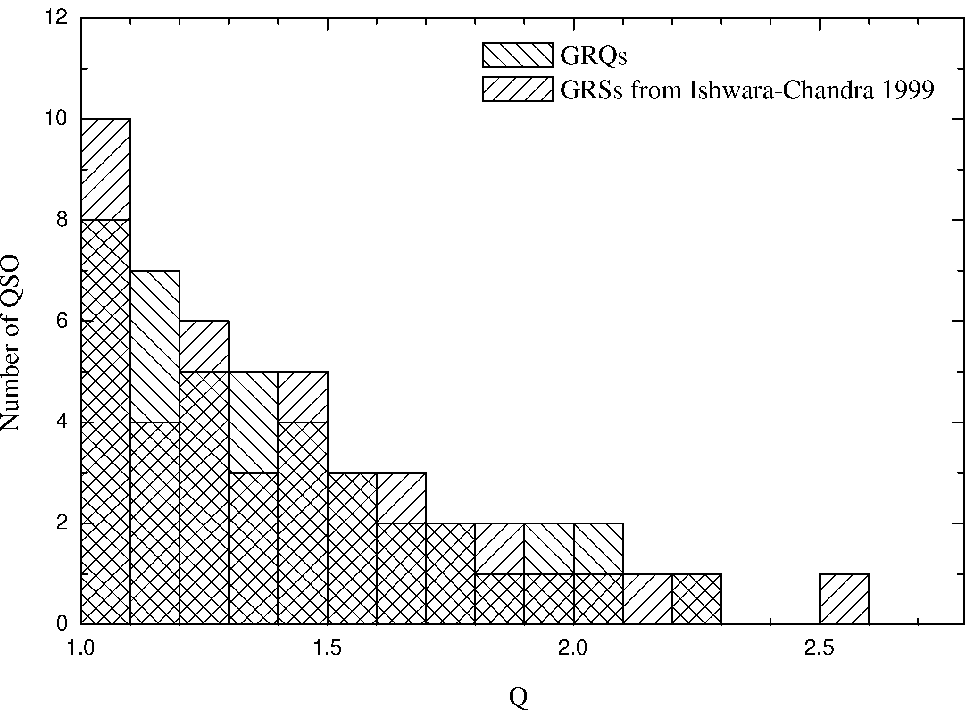}
 \caption{The distributions of the arm-length-ratio parameter $Q$. The top diagram shows all radio quasars from our samples, while the bottom one includes GRQs from our sample and GRSs taken from \citealt{b25}. The observed distribution of the Q parameter suggests that the IGM in which the giants evolve is not more symmetrical than that around the smaller sources.}
\end{figure} 

\begin{figure}
 \includegraphics[width=0.99\linewidth]{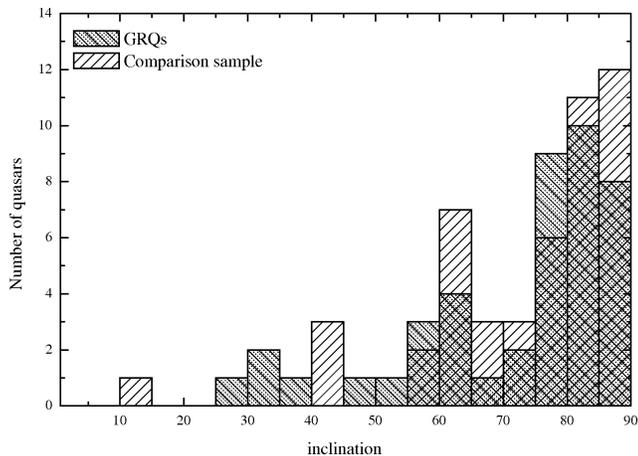}
 \caption{Distribution of the inclination angle, $i$, for the samples of radio quasars. For the definition of the inclination angle see Sect.~4.1. $i$=90\degr means that the jets and lobes lie in the plane of the sky.}
\end{figure}

\subsection[]{Black hole mass estimations}
In order to obtain the central BH mass of quasars from our samples we used
measurements of the CIV, MgII and H$\beta$ emission lines and the mass-scaling relations (equations (6), (7) and (8)). The mass values obtained are in the range of ~ $1.6 \cdot 10^8 M_{\odot} < M_{\rm BH} < 12.3 \cdot 10^8 M_{\odot}$ when using the MgII emission line, and $1.5 \cdot 10^8 M_{\odot} < M_{\rm BH} < 29.2 \cdot 10^8 M_{\odot}$ when using the $H_{\beta}$ emission line. For some GRQs and quasars from the comparison sample it was possible to compare the results obtained on the basis of different emission-line measurements. In Fig.~10 we present the relation between the mass values calculated from MgII vs $H_{\beta}$ lines and those from CIV vs MgII lines, respectively. We found that the mass estimations based on the MgII line on average tend to be smaller than those obtained using the $H_{\beta}$ emission line (the linear fit to the data points is given by the relation: $M_{BH}H_{\beta}=2.87(\pm 0.98)\cdot M_{BH}MgII+5.00(\pm 9.25)$), and the mass estimations based on CIV line are larger than those obtained from the MgII line ($M_{BH}CIV=0.68(\pm 0.14) \cdot M_{BH}MgII+1.08(\pm 0.61)$). The above results are consistent with the earlier comparisons of BH masses estimated by other authors (e.g. \citealt{b63, b16, b62}).

\begin{figure}
\centering
 \includegraphics[width=0.99\linewidth]{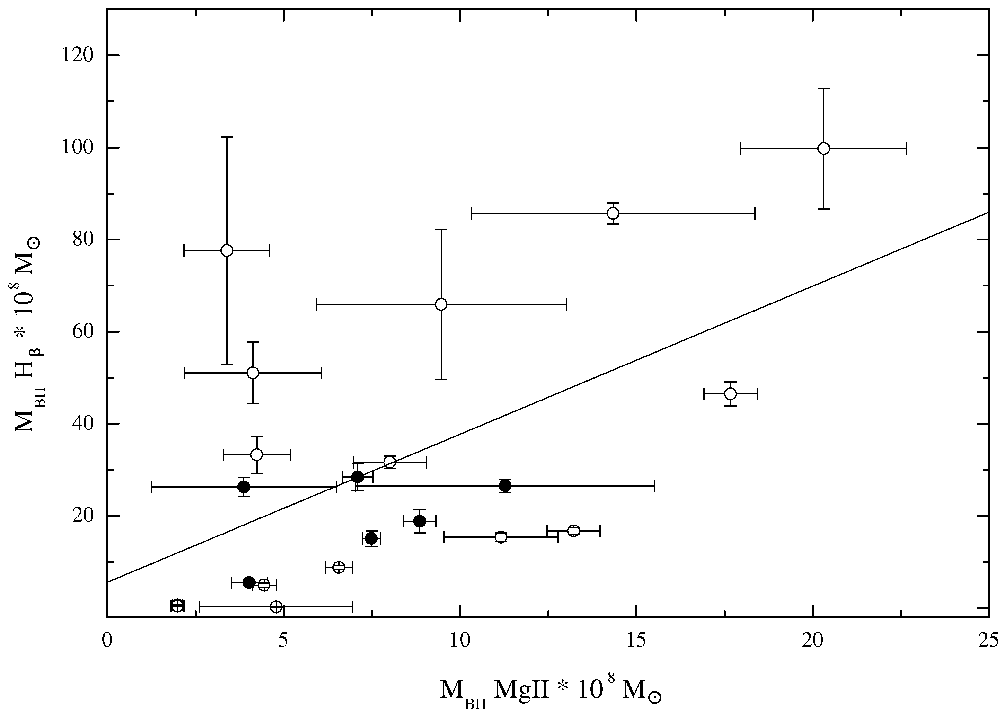}\\
 \hspace{0.5cm}\includegraphics[width=0.99\linewidth]{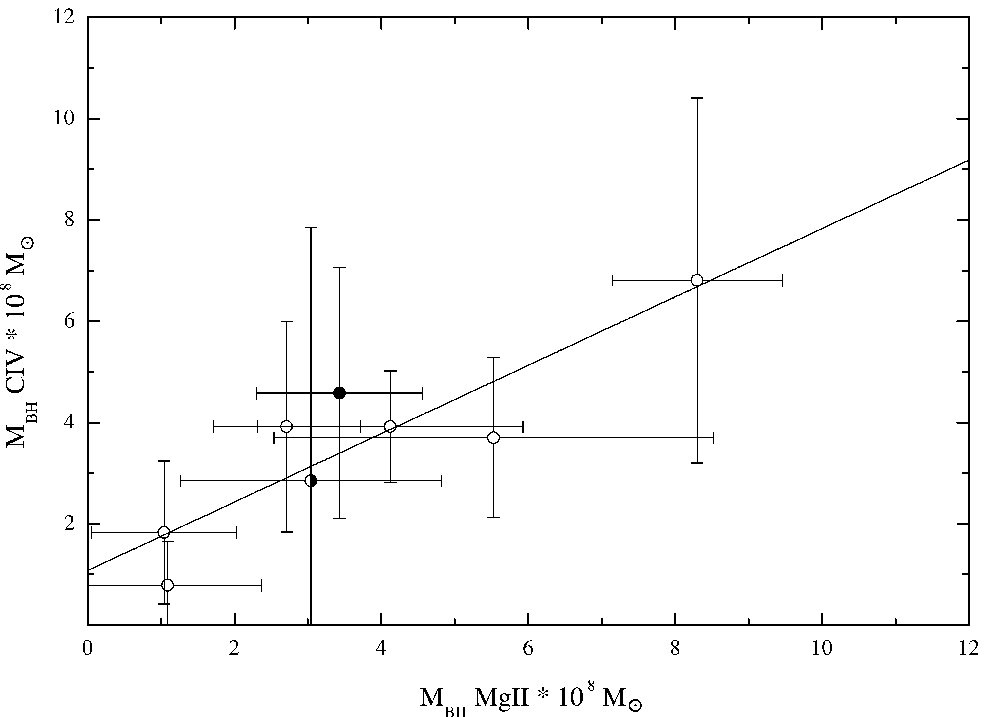}
 \caption{Comparison of BH mass values estimated using measurement of different emission lines.  {\bf top diagram:} MgII versus $H_{\beta}$ BH masses; {\bf bottom diagram:} MgII versus CIV BH mass. The linear fits to data points are described in the text.}
\end{figure}


\subsection[]{Black hole mass vs radio properties}
In the paper by \cite{bkp}, it is claimed that in the jet-formation models some dependence of jet power on BH mass should be expected. The assumption that the giants are formed due to a longer activity phase of the central AGN and/or more frequent duty cycles can imply that their BH masses should be larger because of longer accretion episodes. In Fig.~11 we present the relations  between the total and core radio luminosity and the BH masse.  It can be distinctly seen, however, that there is no correlation between the BH mass and either the core luminosity or the total luminosity  for GRQs as well as smaller radio quasars.

\begin{figure}
\centering
 \includegraphics[width=0.99\linewidth]{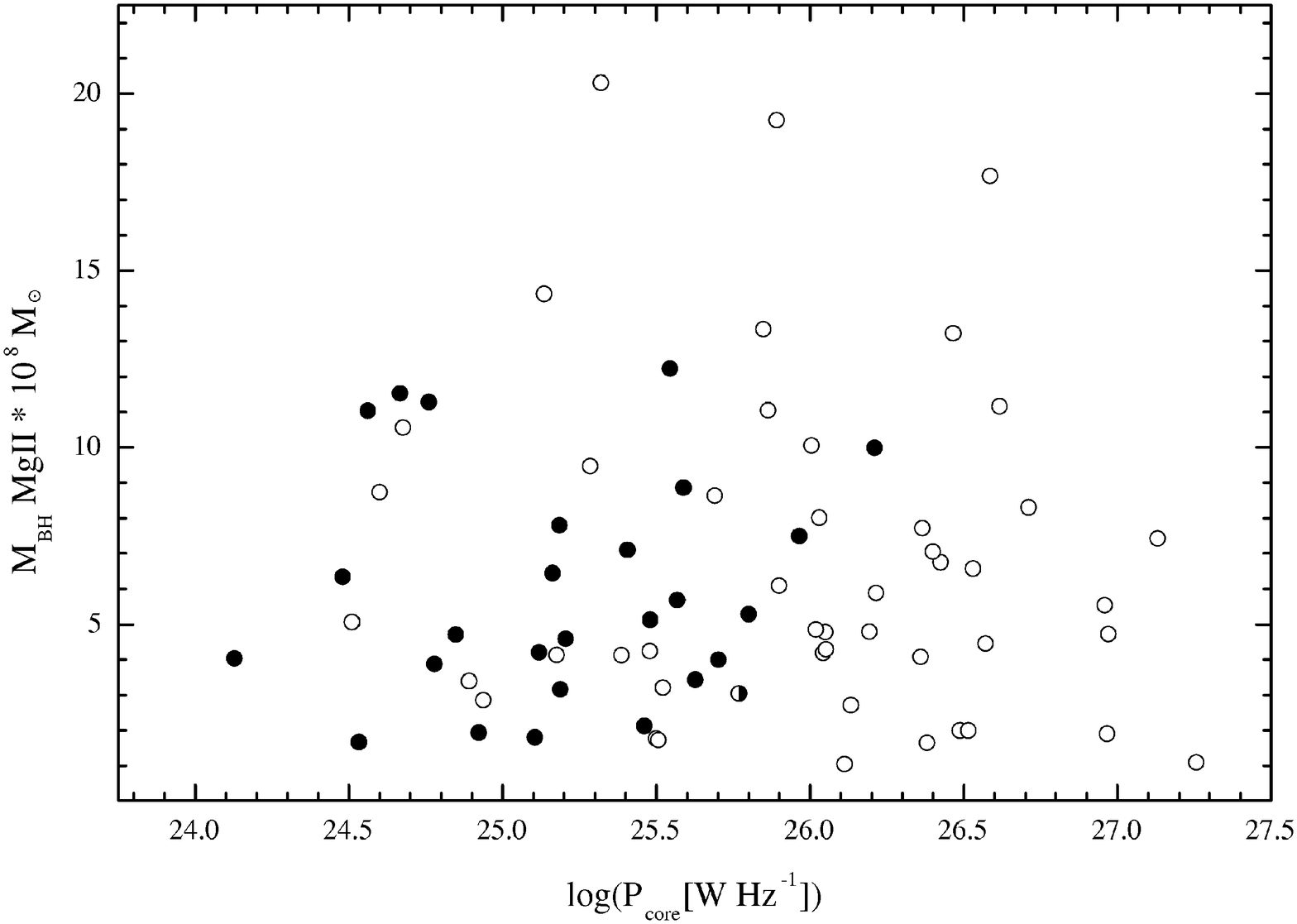}\\
 \hspace{1cm}\includegraphics[width=0.99\linewidth]{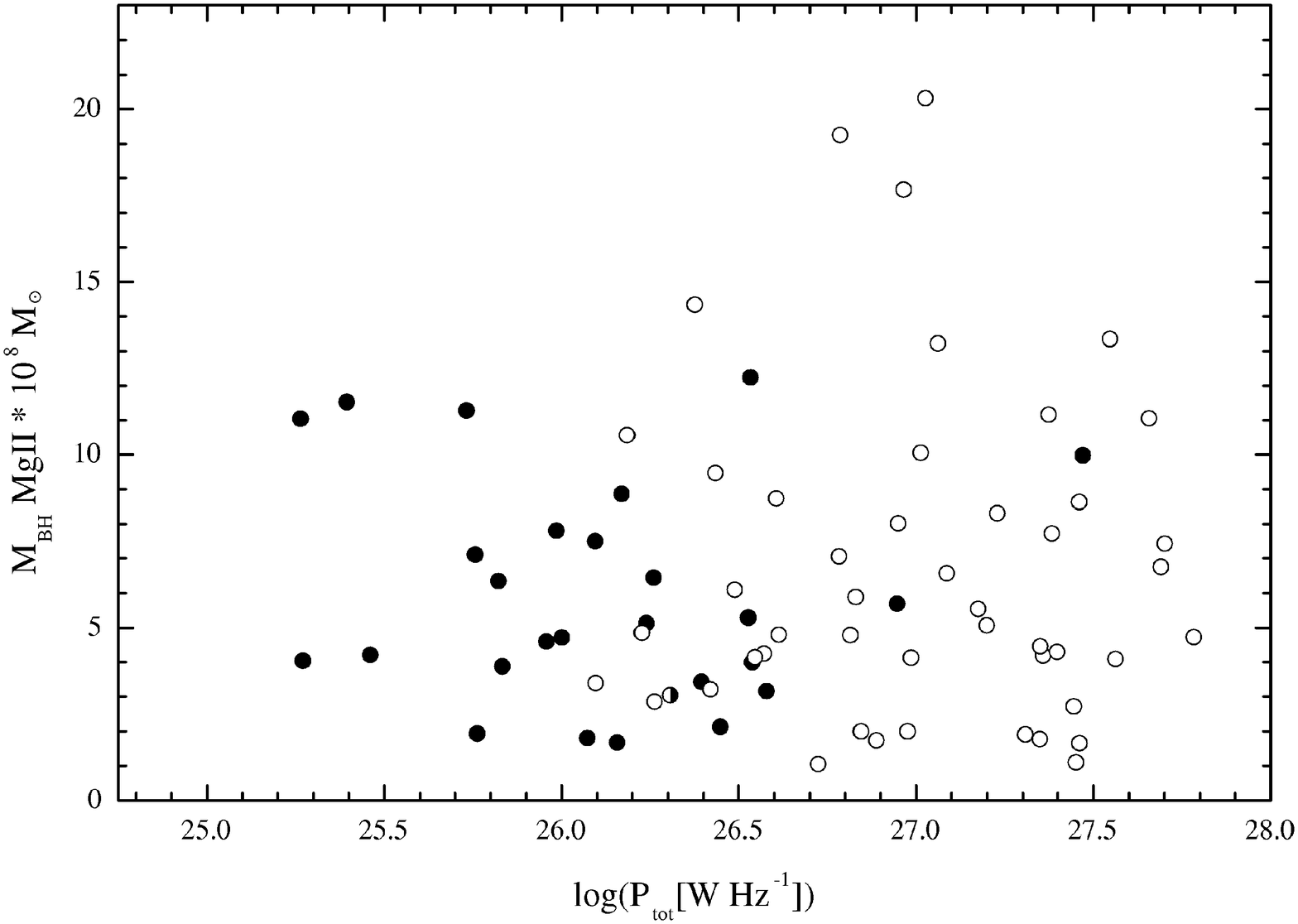}
 \caption{Relations between BH mass and radio luminosity at 1.4 GHz. {\bf top diagram:} BH mass vs core luminosity; {\bf bottom diagram:} BH mass vs total luminosity.}
\end{figure}

We also looked for a relation between BH masse and the unprojected radio linear size of a radio quasars. For the MgII BH mass estimations (Fig.~12), no obvious dependence has been found. Some interesting results can, however, be seen in Fig.~13. For the H$\beta$ and CIV BH mass estimations it can be clearly observed that the dependence between linear size of radio structures  and their BH mass is quite significant. Surprisingly enough, the relation based on the H$\beta$ mass estimations for GRQs does not at all resemble that for quasars from the comparison sample. The slope of the linear fit for the sample of smaller quasars is steeper than that for the GRQs sample. This result suggests that the GRQs can be considered to represent another group of objects which differ physically from smaller quasars. We fitted linear functions independently to the data of GRQs and to the comparison sample (left panel of Fig.~13). The best fits obtained are as follows: $M_{\rm BH}H\beta = 10.995(\pm 7.023) \cdot D^*+1.629(\pm 10.268)$ and $M_{\rm BH}H\beta = 95.830(\pm 25.392) \cdot D^*-5.996(\pm 12.011)$ for the GRQs and for the comparison sample, with correlation coefficients of 0.48 and 0.74, respectively. We also plotted these lines on Fig.~12, taking into account the scaling factor between H$\beta$ and MgII BH mass estimations (equal to 2.87). It is obvious that the giants and the smaller radio quasars fulfil these relations quite well. Moreover, for the CIV mass estimation a weak correlation is also observed. The best fit is represented by a line $M_{\rm BH}CIV = 9.720(\pm 4.589) \cdot D^*+1.620(\pm 3.276)$ with a correlation coefficient of 0.51. The result obtained (particularly for the $H\beta$ mass estimations) can indicate that there may be some difference between GRQs and smaller radio quasars. It is hard to find a physical process to account for such a behaviour, especially as it is not found in the diagram for CIV BH masse. Some authors (e.g. \citealt{b30, b15}) suggested that the formation of different emission lines occurred in different regions of the broad line region, in the sense that the CIV emission should originate below the $H\beta$ emission. Therefore, GRQs and smaller radio quasars may differ with respect to the external structures of the broad line region, while their central parts would be similar. The question now is how to reconcile the fact that, according to the previously analysed relations for GRQs and smaller quasars, we did not see any clear distinction between these two types and here there is a clear difference. The possibility which comes to mind is that there is a difference in age between GRQs and smaller quasars and the composition of the broad-line region could be different for young and old quasars.  However, the number of sources analysed using the CIV mass estimations is too small to allow for any definite conclusions, particularly relating to the smaller radio quasars. For example, this correlation deteriorates if we artificially shift the defining minimum GRQ size from 0.72 Mpc to a smaller value.
Generally, apart from the above speculations on the composition of the broad line region, we can conclude that the apparent relationship between the linear size of the radio structure and the BH mass supports the evolutionary origin of GRQs: as time increases, the BH mass becomes larger and the size of radio structure grows.

\begin{figure}
\centering
 \includegraphics[width=0.99\linewidth]{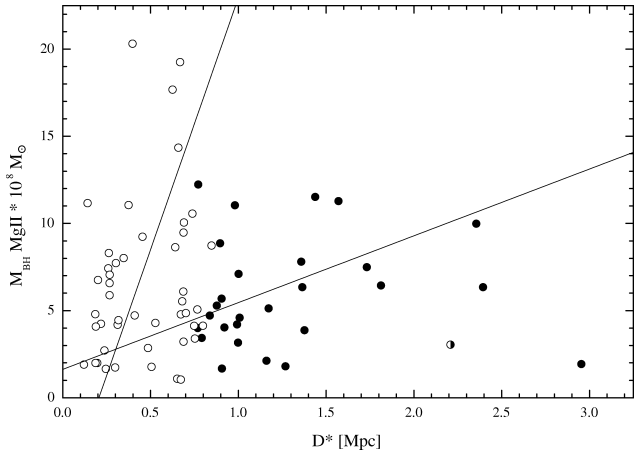}
 \caption{Dependence between the BH masses derived from the MgII emission line, and the unprojected linear sizes of the radio structures. The straight lines are reproduced from Fig.~13 (for details see the text).}
\end{figure}

\begin{figure}
 \hspace{0.3cm}\includegraphics[width=0.99\linewidth]{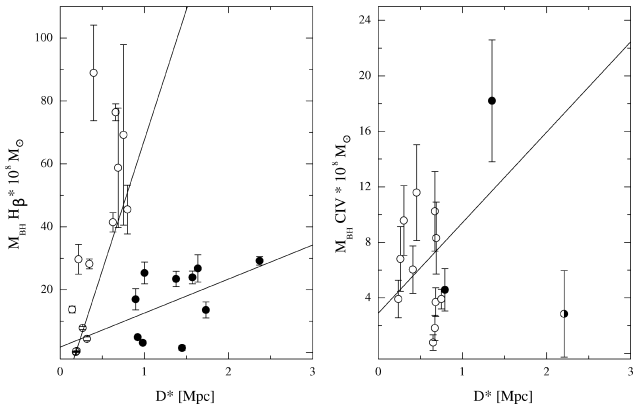}
 \caption{Dependence between the BH mass derived from the H$\beta$ - {\bf left panel}, and the CIV - {\bf right panel} - emission line, and the unprojected linear size of the radio structure.}
\end{figure}

\subsection[]{Accretion rate}
Using the obtained BH mass and the optical monochromatic continuum luminosity ($\lambda L_{\lambda}$) we calculated the accretion rate for our sample of quasars. The accretion rate is computed as \.m$(\lambda)$ = $L_{\rm bol}/L_{\rm Edd}$, where $L_{\rm bol}$ is the bolometric luminosity, assumed as:

    \begin{equation}
    L_{bol}=C_{\lambda} \lambda L_{\lambda}\label{eq10}
    \end{equation}

\noindent
where $C_{\lambda}$ is equal: 9.0 for ${\lambda}=5100\rm\AA$ (according to \citealt{b28}), 5.9 for ${\lambda}=3000\rm\AA$ (according to \citealt{b45}) and  4.6 for ${\lambda}=1350\rm\AA$ (according to \citealt{b61}). Following \cite{b16} the Eddington luminosity $L_{\rm Edd}$ is given by:

\begin{equation}
    L_{\rm Edd}=1.45\cdot10^{38} M_{\rm BH}/M_{\odot} erg s^{-1}
\label{eq11}
\end{equation}

The resulting values of $L_{\rm bol}$, $L_{\rm Edd}$ and \.m$(\lambda)$ for GRQs and smaller quasars are listed in Table 6 and 7, respectively. In Fig.~14 we present the BH mass as a function of accretion-rate values, which are calculated basing on the CIV, MgII and H$\beta$ emission lines as well as on the respective continuum luminosities, taking into account the scaling factor between H$\beta$, CIV and MgII mass estimations. As can be seen, the accretion rate is apparently higher for less massive BHs. A similar result was obtained by \cite{b16} for a sample of quasars and by \cite{b40} for narrow-line Seyfert galaxies. The result is consistent with the scenario of quasars increasing their BH mass during the accretion process solely. When there is no matter left, the accretion rate decreases, while a large amount of mass could have been accumulated in the central BH during the previous accretion episodes. In the scenario described by \cite{b40}, the accretion rate is high in the early stages of AGN evolution and drops later on, so we could expect that at higher redshifts we should observe objects with larger accretion rates. However, Fig.~15 shows that, for our samples of quasars, no dependence between accretion rate and redshift is seen.\\
The accretion rates for GRQs and for the comparison sample are consistent with typical values (0.01 $\div$ 1) for AGNs. Given the observed accretion rate we can constrain the lifetimes of the BHs in our samples. The obtained lower value for GRQs imply, that these sources are more evolved systems, for which the e-folding time to increase their BH mass (for a definition see e.g. \citealt{b2223}) is longer than in the case of smaller-size quasars. The obtained mean values of accretion rate (\.m(3000\AA)) are  $0.07 \pm 0.03$ and $0.09 \pm 0.07$, respectively, for GRGs and smaller-size radio quasar. The dependence between accretion rate and unprojected linear size of radio structure is presented in Fig.~16. 

\begin{figure}
 \includegraphics[width=0.99\linewidth]{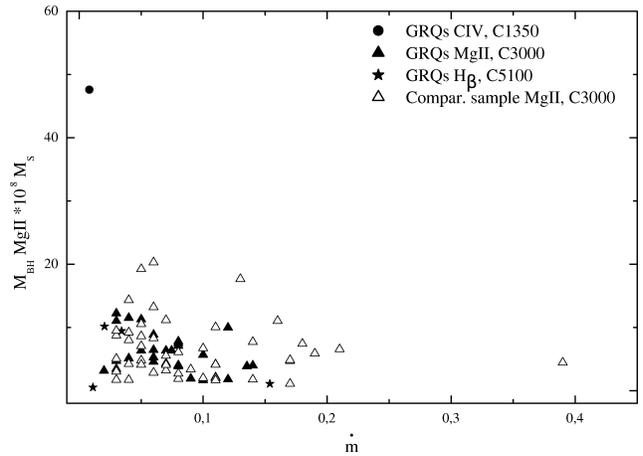}
 \caption{The dependence between BH mass and accretion rate \.m$(\lambda)$. The solid and open symbols mark GRQs and smaller-size radio quasar, respectively. Different symbols (circles, triangles and stars) represent estimations of accretion rate base on measurement of different emission lines (MgII, CIV and H$\beta$) and luminosities (at $\lambda=1350\rm\AA$,   
$\lambda=3000\rm\AA$ or $\lambda=5100\rm\AA$).}
\end{figure}
\begin{figure}
 \includegraphics[width=0.99\linewidth]{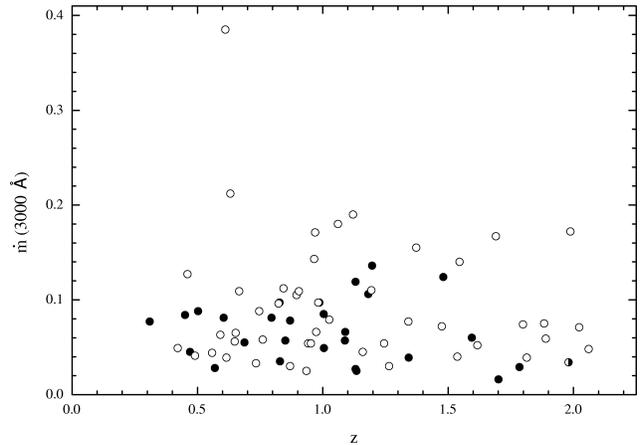}
 \caption{BH accretion rate vs redshift. No correlation is seen.}
\end{figure}
In Fig.~17 we present the dependence of accretion rate \.m(3000\AA) onto the core, as well as total, radio luminosity. There is a distinct trend for larger accretion rates to be observed in quasars with larger radio luminosity. The linear fits for \.m$(3000\rm\AA)$ are described by:\\
\.m(3000\AA)=0.114($\pm$0.044)log($P_{tot}$)-4.193($\pm$1.185),\\
\.m(3000\AA)=0.138($\pm$0.038)log$(P_{core})$-4.696($\pm$0.984)\\
with correlation coefficients equal to 0.29 and 0.39 respectively. 

\begin{figure}
\centering
 \includegraphics[width=0.99\linewidth]{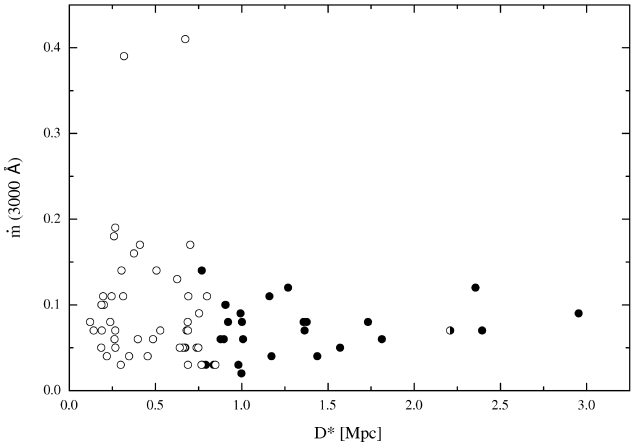}\\
 \caption{Accretion rate vs unprojected linear size of radio structure.}
\end{figure}

\begin{figure}
 \includegraphics[width=0.99\linewidth]{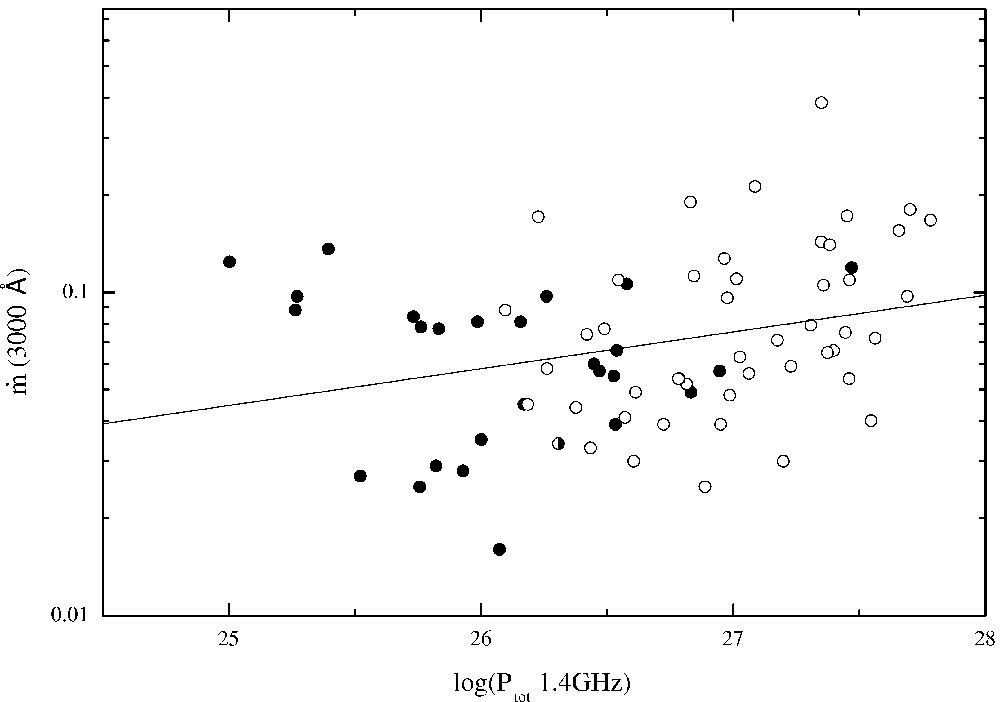}\\
\hspace{0.5cm}\includegraphics[width=0.99\linewidth]{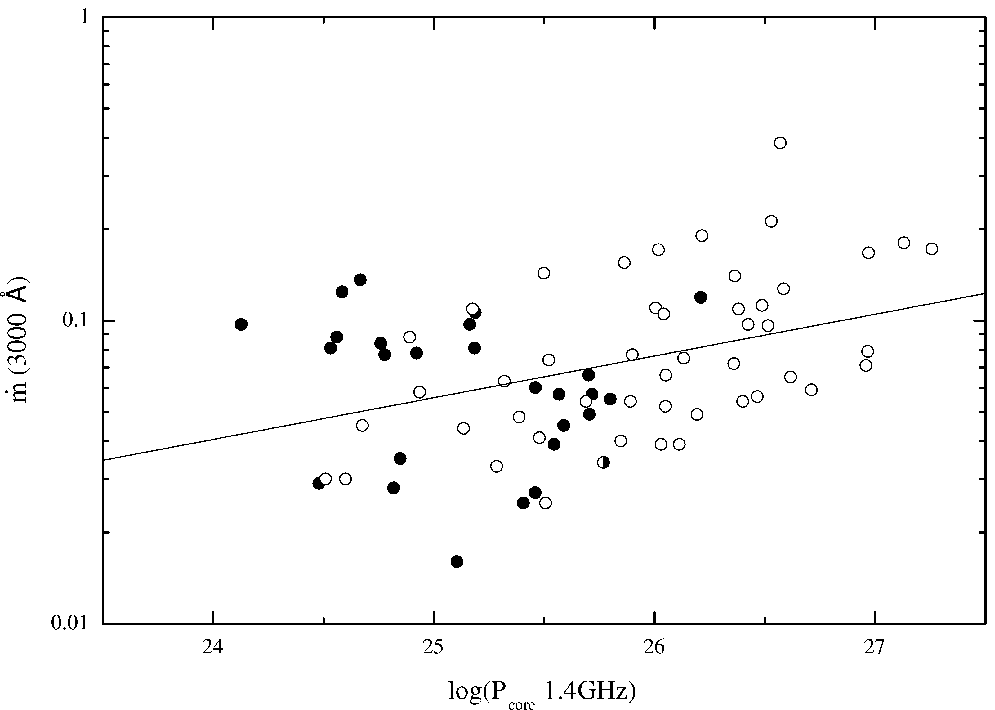}
 \caption{Accretion rate as a function of total radio luminosity - {\bf top panel} and core radio luminosity - {\bf bottom panel}. }
\end{figure}


\section[]{Conclusions}
We have presented a comparison of radio and optical properties for a sample of GRQs and smaller radio quasars. It is important to mention that the measurements were obtained in a similar, homogeneous, manner for all sources from both the GRQ and comparison samples. Only the absolute values may be affected by some global calibration errors, if at all. The final conclusions are summarized below: \\

\begin{enumerate}

\item  
Based on the $P$--$D$ diagram, we found that there is a continuous distribution of GRQs and smaller radio quasars. Therefore we can conclude that the GRQs could have evolved over time out of smaller radio quasars, which is consistent with the predictions of evolutionary models. We did not find that GRQs should have more prominent radio cores, which could suggest that the giants are similar to the smaller objects if we take into account their radio energetics.\\

\item 
The arm-length-ratio and bending angle values for both GRQs and smaller radio quasars are similar, which indicates that there is no significant difference of the environmental properties of the IGM within which giant- and smaller radio quasars evolve.\\

\item 
Statistically, the inclination angles obtained for our samples of quasars are inconsistent with traditional AGN unification scheme. Inclinations larger than 45$^{ \rm o}$ could, however, be explained based on recent results from studies of dusty torus properties.\\

\item 
The  values of BH masses estimated here are similar to those for the powerful AGNs. The BH masses estimated using the MgII emission line are in the range of $1.6 \cdot 10^8 M_{\odot} < M_{\rm BH} < 12.2 \cdot 10^8 M_{\odot}$  and $1.0 \cdot 10^8 M_{\odot} < M_{\rm BH} < 20.3 \cdot 10^8 M_{\odot}$ for GRQs and for the smaller radio quasars respectively. We did not find any constraints for more massive BHs to be located in GRQs.\\

\item 
We did not find any significant correlation between the BH mass and the radio luminosity. However, using the $H_{\beta}$ and CIV line BH mass estimations a weak correlation between the linear size of the radio structure and the BH mass has been revealed. This might suggest that the linear size of giants could be related to their ``central engines''. Surprisingly enough, the same relation, but based on the $H_{\beta}$ analysis results, is different for the GRQs and for the smaller radio quasars, which could suggest an inherent difference between these types of objects. However, this result should be taken with some caution as it was obtained only for a small number of quasars. The relation between the linear size of the radio structure and the BH mass supports the evolutionary origin of GRQs.\\

\item 
The accretion rate for the more massive BHs is smaller than that for the less massive BHs. It is consistent with the scenario that quasars increase their BH mass during accretion process. The obtained mean value of accretion rate is equal to 0.07 for GRQs and 0.09 for smaller radio quasars. The lower value for GRQs suggests that GRQs are more evolved (aged) sources whose accretion process has slowed down or is almost over.
The difference of \.m($\lambda$) and BH mass between the  small-size radio quasars and large-size ones is, however, not significant, which could indicate similarities in their evolution. We found also a weak correlation between the accretion rates and the core radio luminosity, which confirms a connection between the accretion processes and the radio emission.\\

\item 
The results obtained from the measurements based on the H$\beta$ and CIV emission lines seem to be more homogeneous than those based on MgII. The BH masses derived from the H$\beta$ and CIV mass-scaling relations have smaller uncertainties than those of the MgII line. The large uncertainties in the case of MgII measurements are due to the fact that this line is strongly affected by the Fe emission. Moreover, the large uncertainty of the mass-scaling relation slope for the MgII line is also due to the absence of reverberation data from systematic monitoring.\\

\end{enumerate}

In summary, taking into account the optical and radio properties, we can conclude that except for their size, the GRQs are similar to the smaller radio quasars. Their BH mass, accretion rate, prominence of radio core are comparable. The environment properties of GRQs and smaller radio quasars are also similar. Therefore, GRQs could be just evolved (aged) radio sources in which the accretion process has been diminished or is almost over and the large size is the consequence of their evolution. The sample of GRQs presented here, which is the largest one known to date, can be used for other astrophysical studies, such as on the evolution of radio sources.

\section*{Acknowledgments}
We are grateful to Richard White for providing us with a number of quasar spectra and Marianne Vestergaard for the template of the Fe emission. We thank J. Machalski, S. Zo{\l}a and D. Kozie{\l}-Wierzbowska for their detailed and very helpful comments on the manuscript. We thank also the anonymous referee for her/his very valuable comments. This project was supported in part by the Polish National Center of Science under decision DEC-2011/01/N/ST9/00726.

\begin{table*}
 \centering
 \begin{minipage}{180mm}
  \caption{Parameters of radio structure for GRQs and smaller-size radio quasars from the comparison sample.}
  \begin{tabular}{lcccccc|lcccccc}
  \hline
IAU              &log(P$_{tot})$&log(P$_{core}$)& B&Q&F   &i     &IAU           &log(P$_{tot})$ &log(P$_{core}$)&B&Q  & F  &i \\
name                 &W/Hz    &W/Hz &[$^o$]&     &     &[$^o$]&name             &W/Hz       &W/Hz  &[$^o$] &     &     & [$^o$]\\
(1)                  & (2)    & (3) &(4)   &(5)  & (6) & (7)  &(1)              & (2)       & (3)  &(4)   &(5)  & (6) & (7)\\
\hline
\multicolumn{7}{l}{Giant quasars}                            
&\multicolumn{7}{l}{Comparison sample}\\
J0204$-$0944&	25.76& 24.92& 0.0 & 2.06& 0.59& 81 &J0022$-$0145& 26.62& 25.13	& 6.7 & 1.07& 2.08& 71\\
J0210$+$0118&	25.99& 25.31& 25.6& 1.38& 0.30& 63 &J0034$+$0118& 27.20& 24.51	& 7.9 & 1.90& 0.26& 60\\ 
J0313$-$0631&	25.89& 24.45& 5.6 & 1.11& 0.97& 87 &J0051$-$0902& 26.61& 24.60	& 9.1 & 1.39& 5.35& 55\\
J0439$-$2422&	27.09& $-$  & 4.5 & 1.67& 0.55& 79 &J0130$-$0135& 26.18& 24.68	& 5.9 & 1.90& 0.15& 62\\
J0750$+$6541&	26.39& 25.73& 5.5 & 1.05& 0.33& 65 &J0245$+$0108& 27.55& 25.85   & 12.0& 2.28& 0.68 & 85\\
J0754$+$3033&	26.10& 25.97& 18.8& 1.77& 1.20& 87 &J0745$+$3142& 26.96& 26.59 	& 5.3 & 1.10& 0.62 & 88\\
J0754$+$4316&	25.63& 24.68& 0.2 & 1.07& 0.36& 85 &J0811$+$2845& 27.23& 26.71 	& 6.1 & 2.35& 1.26 & 80\\
J0801$+$4736&	25.00& 24.58& 8.7 & 1.05& 2.21& 37 &J0814$+$3237& 26.84& 26.49 	& 12.8& 2.40& 0.37 & 72\\
J0809$+$2912&	27.47& 26.21& 1.5 & 1.25& 0.04& 28 &J0817$+$2237& 27.69& 26.42 	& 5.7 & 1.06& 2.38 & 71\\
J0812$+$3031&	26.07& 25.10& 2.4 & 1.37& 2.91& 71 &J0828$+$3935& 26.26& 24.94 	& 0.4 & 1.15& 0.23 & 79\\
J0819$+$0549&	26.58& 25.19& 0.0 & 1.24& 1.54& 81 &J0839$+$1921& 27.78& 26.97 	& 11.4& 1.36& 0.11 & 41\\
J0842$+$2147&	26.45& 25.46& 0.9 & 2.28& 0.92& 85 &J0904$+$2819& 26.83& 26.22 	& 2.3 & 1.07& 3.15 & 45\\
J0902$+$5707&	26.53& 25.80& 9.3 & 1.38& 2.74& 79 &J0906$+$0832& 26.81& 26.05	& 0.5 & 1.29& 0.88 & 86\\
J0918$+$2325&	26.17& 25.59& 9.4 & 1.48& 0.99& 81 &J0924$+$3547& 26.49& 25.90	& 1.9 & 1.05& 0.68 & 84\\
J0925$+$4004&	25.73& 24.76& 5.7 & 1.13& 1.18& 80 &J0925$+$1444& 27.36& 26.04 	& 5.6 & 1.26& 1.25 & 84\\
J0937$+$2937&	25.27& 24.13& 4.3 & 1.56& 0.69& 81 &J0935$+$0204& 27.06& 26.47 	& 4.3 & 1.62& 37.20& -\\
J0944$+$2331&	26.95& 25.57& 7.1 & 1.67& 0.44& 83 &J0941$+$3853& 26.95& 26.03 	& 0.3 & 1.44& 0.72 & 85\\
J0959$+$1216&	25.96& 25.21& 11.0& 1.17& 1.80& 73 &J0952$+$2352& 26.23& 26.02	& 17.7& 2.56& 0.97 & 89\\
J1012$+$4229&	25.52& 25.46& 15.6& 1.39& 3.71& 59 &J1000$+$0005& 27.46& 26.38 	& 18.4& 1.35& 1.17 & 85\\
J1020$+$0447&	26.00& 27.85& 6.7 & 1.08& 3.71& 61 &J1004$+$2225& 27.40& 26.05 	& 4.7 & 1.25& 1.03 & 85\\
J1020$+$3958&	25.39& 24.67& 1.9 & 1.12& 3.62& 58 &J1005$+$5019& 27.18& 26.96	& 16.2& 2.73& 1.54 & 76\\
J1027$-$2312&	26.21& 25.38& 8.2 & 1.15& 0.82& 88 &J1006$+$3236& 27.31& 26.97 	& 0.9 & 2.89& 1.06 & 80\\
J1030$+$5310&	26.54& 25.70& 9.4 & 1.58& 3.22& 77 &J1009$+$0529& 26.79& 25.89	& 8.7 & 1.01& 1.75 & 78\\
J1054$+$4152&	25.82& 24.48& 18.8& 1.18& 4.23& 75 &J1010$+$4132& 27.35& 26.57	& 5.5 & 1.69& 12.73& 42\\
J1056$+$4100&	26.39& 25.63& 7.8 & 2.10& 0.89& 87 &J1023$+$6357& 27.01& 26.00	& 0.2 & 1.37& 0.44 & 70\\
J1130$-$1320&	27.20& 25.40& 0.7 & 1.13& 0.29& 67 &J1100$+$1046& 26.61& 26.19 	& 1.8 & 1.20& 0.61 & 78\\
J1145$-$0033&	26.47& 25.72& 10.2& 1.29& 0.59& 82 &J1100$+$2314& 26.38& 25.14	& $-$  & $-$ & 3.64 & 68\\
J1148$-$0403&	26.32& 25.76& 12.8& 1.06& 1.20& 88 &J1107$+$0547& 26.42& 25.52	& 7.2 & 1.46& 0.63 & 80\\
J1151$+$3355&	26.26& 25.16& 11.3& 1.98& 0.33& 32 &J1107$+$1628& 27.09& 26.53 	& 3.4 & 1.05& 0.67 & 87\\
J1229$+$3555&	26.20& 24.65& 14.9& 1.33& 0.39& 57 &J1110$+$0321& 27.35& 25.50 	& 10.7& 2.38& 0.55 & 85\\
J1304$+$2454&	25.76& 25.41& 1.6 & 1.46& 1.82& 77 &J1118$+$3828& 26.10& 24.89	& 3.5 & 1.25& 3.25 & 55\\
J1321$+$3741&	26.53& 25.55& 17.0& 1.06& 0.72& 79 &J1119$+$3858& 26.43& 25.28	& 8.4 & 1.14& 2.66 & 64\\
J1340$+$4232&	26.24& 25.48& 3.5 & 1.92& 0.59& 89 &J1158$+$6254& 27.03& 25.32 	& 4.3 & 1.53& 0.46 & 75\\
J1353$+$2631&	25.83& 24.78& 13.4& 1.21& 2.98& 34 &J1217$+$1019& 27.44& 26.13 	& 26.8& 1.43& 0.73 & 87\\
J1408$+$3054&	25.93& 24.82& 4.9 & 1.41& 2.72& 80 &J1223$+$3707& 26.57& 25.48 	& 3.9 & 1.75& 0.72 & 83\\
J1410$+$2955&	25.26& 24.56& 10.4& 1.30& 1.00& 81 &J1236$+$1034& 26.55& 25.18 	& 0.4 & 1.42& 0.39 & 63\\
J1427$+$2632&	26.17& 25.23& 9.1 & 1.70& 0.40& 45 &J1256$+$1008& 26.98& 26.51 	& 20.3& 1.14& 0.30 & 69\\
J1432$+$1548&	26.83& 25.71& 2.9 & 1.39& 0.99& 87 &J1319$+$5148& 27.70& 27.13 	& 27.9& 1.81& 0.54 & 61\\
J1504$+$6856&	26.13& 25.52& 4.0 & 1.85& 1.66& 81 &J1334$+$5501& 27.46& 25.69 	& 0.4 & 1.13& 0.79 & 89\\
J1723$+$3417&	26.26& 25.67& 1.1 & 1.05& 2.11& 51 &J1358$+$5752& 27.66& 25.86 	& 1.3 & 1.20& 1.24 & 85 \\
J2042$+$7508&	25.67& 24.72& 7.1 & 1.03& 2.69& 61 &J1425$+$2404& 27.37& 26.62 	& 18.9& 1.41& 1.60 & 83 \\
J2234$-$0224&	25.89& 24.71& 1.9 & 1.49& 0.18& 87 &J1433$+$3209& 26.89& 25.51 	& 13.8& 1.14& 0.96 & 88 \\
J2344$-$0032&	25.46& 25.12& 0.8 & 1.54& 0.76& 79 &J1513$+$1011& 27.38& 26.36 	& 20.9& 1.46& 1.07 & 83 \\
	    &	     &	    &	   &	 &     &   &J1550$+$3652& 26.99& 25.39	& 1.0 & 1.78& 0.28 & 64\\
	    &	     &	    &	   &	 &     &   &J1557$+$0253& 26.78& 26.40	& 7.4 & 3.62& 8.05 & 61\\
	    &	     &	    &	   &	 &     &   &J1557$+$3304& 27.45& 27.26	& 20.2& 1.54& 1.50 & 89 \\
	    &	     &	    &	   &	 &     &   &J1622$+$3531& 27.56& 26.36 	& 17.2& 2.67& 0.59 & 82 \\
	    &	     &	    &	   &	 &     &   &J1623$+$3419& 26.31& 25.77	& 7.9 & 2.19& 0.64 & 13\\
	    &	     &	    &	   &	 &     &   &J2335$-$0927& 26.72& 26.11	& 8.0 & 2.74& 1.66 & 83\\
\hline
\end{tabular}
\end{minipage}
\end{table*}

\begin{table*}
 \centering
 \begin{minipage}{195mm}
  \caption{Parameters of optical spectra and black hole mass for GRQs.}
  \begin{tabular}{@{}lcccccccccccc@{}}
  \hline							
IAU         &       & FWHM  &          &      &f$_\lambda$&   &  &Log$\lambda L_{\lambda}$& & &$M_{BH}$& \\		
name        &   CIV  & MgII  & H$_\beta$& 1350\AA     &3000\AA    &  5100\AA    &1350\AA                  & 3000\AA                &5100\AA                 &  CIV     &  MgII  & H$\beta$ \\			
            & \multicolumn{3}{c|}{\rule[0mm]{4cm}{0.1pt}}& \multicolumn{3}{c|}{\rule[0mm]{3.5cm}{0.1pt}}  & \multicolumn{3}{c|}{\rule[0mm]{3.9cm}{0.1pt}}   & \multicolumn{3}{c|}{\rule[0mm]{4cm}{0.1pt}}\\
            & \multicolumn{3}{c}{\AA}                  & \multicolumn{3}{c}{10$^{-17}$ erg cm$^{-2}$ s$^{-1}$ $\AA^{-1}$}   & \multicolumn{3}{c|}{erg s$^{-1}$ } & \multicolumn{3}{c|}{10$^8$ M$\odot$} \\
\hline
J0204$-$0944&	$-$  & 34.19 &	$-$   &	$-$  &10.13  &\hspace{-0.5cm}$-$   & \hspace{-0.5cm}$-$&   44.61 & $-$      & \hspace{-0.5cm}$-$ 	& 1.93$^{\pm0.25}$ & $-$\\
J0210$+$0118&	$-$  & 49.61 &	$-$   &	$-$  &46.57  &\hspace{-0.5cm}$-$   & \hspace{-0.5cm}$-$&   45.17 & $-$      & \hspace{-0.5cm}$-$ 	& 7.80$^{\pm0.22}$ & $-$\\
J0754$+$3033&	$-$  & 48.59 &	139.90&	$-$  &53.77  &\hspace{-0.5cm}10.65 & \hspace{-0.5cm}$-$&   45.17 & 44.70    & \hspace{-0.5cm}$-$ 	& 7.49$^{\pm0.25}$ & 13.56$^{\pm2.57}$\\
J0754$+$4316&	$-$  & $-$   &	226.37&	$-$  &$-$    &\hspace{-0.5cm}30.2  & \hspace{-0.5cm}$-$&   $-$   & 44.53    & \hspace{-0.5cm}$-$ 	& $-$              & 29.17$^{\pm1.32}$\\
J0801$+$4736&	$-$  & $-$   &	125.23&	$-$  &$-$    &\hspace{-0.5cm}3.69  & \hspace{-0.5cm}$-$&   $-$   & 42.97    & \hspace{-0.5cm}$-$ 	& $-$              & 1.48$^{\pm0.76}$ \\
J0809$+$2912&	$-$  & 46.91 &	$-$   &	$-$  &44.45  &\hspace{-0.5cm}$-$   & \hspace{-0.5cm}$-$&   45.48 & $-$      & \hspace{-0.5cm}$-$ 	& 9.98$^{\pm0.51}$ & $-$\\
J0812$+$3031&	$-$  & 30.91 &	$-$   &	$-$  &11.14  &\hspace{-0.5cm}$-$   & \hspace{-0.5cm}$-$&   44.72 & $-$      & \hspace{-0.5cm}$-$ 	& 1.80$^{\pm0.15}$ & $-$\\
J0819$+$0549&	75.57& 58.94 &	$-$   &	$-$  &1.50   &\hspace{-0.5cm}$-$   & \hspace{-0.5cm}$-$&   44.09 & $-$      & \hspace{-0.5cm}$-$ 	& 3.16$^{\pm2.77}$ & $-$\\
J0842$+$2147&	$-$  & 33.16 &	$-$   &	$-$  &10.96  &\hspace{-0.5cm}$-$   & \hspace{-0.5cm}$-$&   44.74 & $-$      & \hspace{-0.5cm}$-$ 	& 2.12$^{\pm0.48}$ & $-$\\
J0902$+$5707&	31.09& 47.90 &	$-$   &	$-$  &10.40  &\hspace{-0.5cm}$-$   & \hspace{-0.5cm}$-$&   44.89 & $-$      & \hspace{-0.5cm}$-$ 	& 5.28$^{\pm1.28}$ & $-$\\
J0918$+$2325&	$-$  & 55.91 &	173.43&	$-$  &54.28  &\hspace{-0.5cm}8.92  & \hspace{-0.5cm}$-$&   45.08 & 44.52    & \hspace{-0.5cm}$-$ 	& 8.86$^{\pm0.46}$ & 16.95$^{\pm3.37}$\\
J0925$+$4004&	$-$  & 62.30 &	196.95&	$-$  &109.20 &\hspace{-0.5cm}20.40 & \hspace{-0.5cm}$-$&   45.10 & 44.60    & \hspace{-0.5cm}$-$ 	& 11.28$^{\pm4.24}$& 23.90$^{\pm2.01}$\\
J0937$+$2937&	$-$  & 41.23 &	98.86 &	$-$  &78.80  &\hspace{-0.5cm}14.70 & \hspace{-0.5cm}$-$&   44.92 & 44.42    & \hspace{-0.5cm}$-$ 	& 4.03$^{\pm0.51}$ & 4.91$^{\pm0.24}$ \\
J0944$+$2331&	$-$  & 43.36 &	$-$   &	$-$  &34.92  &\hspace{-0.5cm}$-$   & \hspace{-0.5cm}$-$&   45.13 & $-$      & \hspace{-0.5cm}$-$ 	& 5.68$^{\pm0.54}$ & $-$\\
J0959$+$1216&	$-$  & 46.95 &	$-$   &	$-$  &14.32  &\hspace{-0.5cm}$-$   & \hspace{-0.5cm}$-$&   44.81 & $-$      & \hspace{-0.5cm}$-$ 	& 4.59$^{\pm3.26}$ & $-$\\
J1020$+$0447&	$-$  & 57.04 &	$-$   &	$-$  &6.56   &\hspace{-0.5cm}$-$   & \hspace{-0.5cm}$-$&   44.49 & $-$      & \hspace{-0.5cm}$-$ 	& 4.71$^{\pm1.45}$ & $-$\\
J1020$+$3958&	$-$  & 66.70 &	$-$   &	$-$  &33.48  &\hspace{-0.5cm}$-$   & \hspace{-0.5cm}$-$&   45.00 & $-$      & \hspace{-0.5cm}$-$ 	& 11.52$^{\pm5.16}$& $-$\\
J1030$+$5310&	$-$  & 36.46 &	$-$   &	$-$  &26.04  &\hspace{-0.5cm}$-$   & \hspace{-0.5cm}$-$&   45.13 & $-$      & \hspace{-0.5cm}$-$ 	& 3.99$^{\pm0.27}$ & $-$\\
J1054$+$4152&	$-$  & 49.00 &	$-$   &	$-$  &23.03  &\hspace{-0.5cm}$-$   & \hspace{-0.5cm}$-$&   45.01 & $-$      & \hspace{-0.5cm}$-$ 	& 6.34$^{\pm3.40}$ & $-$\\
J1056$+$4100&	34.41& 51.61 &	$-$   &	$-$  &2.841  &\hspace{-0.5cm}$-$   & \hspace{-0.5cm}$-$&   44.39 & $-$      & \hspace{-0.5cm}$-$        & 3.43$^{\pm1.13}$ & $-$\\
J1145$-$0033&	61.12& $-$   &	$-$   &	16.07&5.38   &\hspace{-0.5cm}$-$   & \hspace{-0.5cm}44.87 &44.74 & $-$      & \hspace{-0.5cm}18.42$^{\pm2.43}$& $-$  & $-$\\
J1151$+$3355&	$-$  & 51.11 &	$-$   &	$-$  &29.15  &\hspace{-0.5cm}$-$   & \hspace{-0.5cm}$-$&   44.95 & $-$      & \hspace{-0.5cm}$-$ 	& 6.44$^{\pm2.48}$ & $-$\\
J1229$+$3555&	$-$  & 31.93 &	$-$   &	$-$  &13.52  &\hspace{-0.5cm}$-$   & \hspace{-0.5cm}$-$&   44.60 & $-$      & \hspace{-0.5cm}$-$ 	& 1.67$^{\pm0.29}$ & $-$\\
J1304$+$2454&	$-$  & 47.98 &	189.27&	$-$  &79.83  &\hspace{-0.5cm}$-$   & \hspace{-0.5cm}$-$&   45.15 & $-$      & \hspace{-0.5cm}$-$ 	& 7.10$^{\pm0.43}$ & $-$\\
J1321$+$3741&	$-$  & 73.96 &	$-$   &	$-$  &15.56  &\hspace{-0.5cm}$-$   & \hspace{-0.5cm}$-$&   44.87 & $-$      & \hspace{-0.5cm}$-$ 	& 12.23$^{\pm3.98}$& $-$\\
J1340$+$4232&	$-$  & 52.94 &	$-$   &	$-$  &8.22   &\hspace{-0.5cm}$-$   & \hspace{-0.5cm}$-$&   44.69 & $-$      & \hspace{-0.5cm}$-$ 	& 5.12$^{\pm2.67}$ & $-$\\
J1353$+$2631&	$-$  & 41.75 &	206.40&	$-$  &136.10 &\hspace{-0.5cm}35.48 & \hspace{-0.5cm}$-$&   44.86 & 44.51    & \hspace{-0.5cm}$-$ 	& 3.87$^{\pm2.63}$ & 23.69$^{\pm2.59}$\\
J1410$+$2955&	$-$  & 69.94 &	$-$   &	$-$  &47.28  &\hspace{-0.5cm}$-$   & \hspace{-0.5cm}$-$&   44.88 & $-$      & \hspace{-0.5cm}$-$	&11.04$^{\pm6.67}$& $-$\\
J1427$+$2632&	$-$  & $-$   &	195.90&	$-$  &$-$    &\hspace{-0.5cm}42.12 & \hspace{-0.5cm}$-$&   $-$   & 44.72    & \hspace{-0.5cm}$-$ & $-$	        & 27.08$^{\pm4.74}$\\
J1432$+$1548&	$-$  & 52.83 &	$-$   &	$-$  &18.82  &\hspace{-0.5cm}$-$   & \hspace{-0.5cm}$-$&   44.88 & $-$      & \hspace{-0.5cm}$-$ & 6.34$^{\pm1.02}$ & $-$\\
J1723$+$3417&	$-$  & $-$   &	64.88 &	$-$  &168.40 &\hspace{-0.5cm}139.60& \hspace{-0.5cm}$-$&   44.62 & 44.77    & \hspace{-0.5cm}$-$ 	& $-$	          & 3.16$^{\pm0.407}$ \\
J2344$-$0032&	$-$  & 41.18 &	$-$   &	$-$  &70.72  &\hspace{-0.5cm}$-$   & \hspace{-0.5cm}$-$&   44.96 & $-$      & \hspace{-0.5cm}$-$ 	& 4.20$^{\pm0.31}$ & $-$\\
\hline
\end{tabular}
\end{minipage}
\end{table*}
										
\begin{table*}
 \centering
 \begin{minipage}{195mm}
  \caption{Parameters of optical spectra and black hole mass for smaller-size
radio quasars.}
  \begin{tabular}{@{}lcccccccccccc@{}}
  \hline						
IAU         &    & FWHM  &          & &\hspace{-0.2cm}f$_\lambda$& & &Log$\lambda L_{\lambda}$& & &$M_{BH}$& \\		
name        &   CIV  & MgII  & H$_\beta$               & \hspace{-0cm}1350\AA                  &\hspace{-0.2cm}3000\AA    &\hspace{-0.4cm}5100\AA      &1350\AA                 & 3000\AA                &5100\AA             &  CIV     &  MgII & H$\beta$\\			
            & \multicolumn{3}{c|}{\rule[0mm]{4cm}{0.1pt}}& \multicolumn{3}{c|}{\rule[0mm]{3.5cm}{0.1pt}}  &\multicolumn{3}{c|}{\rule[0mm]{3.9cm}{0.1pt}}  & \multicolumn{3}{c|}{\rule[0mm]{4cm}{0.1pt}}\\
            & \multicolumn{3}{c}{\AA}                  &\multicolumn{3}{c}{10$^{-17}$erg cm$^{-2}$s$^{-1}$$\AA^{-1}$}  &\multicolumn{3}{c|}{erg s$^{-1}$ }                                         & \multicolumn{3}{c|}{10$^8$ M$\odot$} \\
\hline
J0034$+$0118&	$-$  & 56.34&	\hspace{-0.4cm}$-$&	\hspace{-0.4cm}$-$& \hspace{-0.4cm}11.80 &	\hspace{-0.05cm}$-$&	\hspace{-0.5cm}$-$   &\hspace{-0.5cm}44.58 &	\hspace{-0.5cm}$-$   &	\hspace{-0.5cm}$-$ &	\hspace{-0.05cm}5.06$^{\pm0.11}$ &	$-$\\
J0051$-$0902&	$-$  & 64.84&	\hspace{-0.4cm}$-$&	\hspace{-0.4cm}$-$& \hspace{-0.4cm}11.51 &	\hspace{-0.05cm}$-$&	\hspace{-0.5cm}$-$   & \hspace{-0.5cm}44.81 &	\hspace{-0.5cm}$-$   &	\hspace{-0.5cm}$-$ &	\hspace{-0.05cm}8.73$^{\pm3.13}$ &	$-$\\
J0130$-$0135&	$-$  & 61.47&	\hspace{-0.4cm}$-$&	\hspace{-0.4cm}$-$& \hspace{-0.4cm}23.58 &	\hspace{-0.05cm}$-$&	\hspace{-0.5cm}$-$   & \hspace{-0.5cm}45.06 &	\hspace{-0.5cm}$-$   &	\hspace{-0.5cm}$-$ &	 \hspace{-0.05cm}10.56$^{\pm0.48}$ &	$-$\\
J0245$+$0108&	35.54& 61.03&	\hspace{-0.4cm}$-$&	\hspace{-0.4cm}$-$& \hspace{-0.4cm}12.67 &	\hspace{-0.05cm}$-$&	\hspace{-0.5cm}$-$   & \hspace{-0.5cm}44.96 &	\hspace{-0.5cm}$-$   &	\hspace{-0.5cm}$-$ &	 \hspace{-0.05cm}9.23$^{\pm3.07}$ &	$-$\\
J0745$+$3142&	186.40&53.84&	\hspace{-0.4cm}173.80&	\hspace{-0.4cm}$-$&  \hspace{-0.4cm}499.50&	\hspace{-0.05cm}107.50&	\hspace{-0.5cm}$-$   & \hspace{-0.5cm}45.74 &	\hspace{-0.5cm}45.30 &	\hspace{-0.5cm}$-$ &	 \hspace{-0.05cm}17.67$^{\pm0.76}$&	41.92$^{\pm4.52}$\\
J0811$+$2845&	27.10& 54.05&	\hspace{-0.4cm}$-$&\hspace{-0.4cm}59.34&\hspace{-0.4cm}12.92 &	\hspace{-0.05cm}$-$&\hspace{-0.5cm}45.40 & \hspace{-0.5cm}45.08 &	\hspace{-0.5cm}$-$   &\hspace{-0.5cm}6.88$^{\pm0.88}$&\hspace{-0.05cm}8.30$^{\pm1.16}$&	$-$\\
J0814$+$3237&	$-$  & 32.17&	\hspace{-0.4cm}$-$&	\hspace{-0.4cm}$-$& \hspace{-0.4cm}17.93 &	\hspace{-0.05cm}$-$&	\hspace{-0.5cm}$-$   &\hspace{-0.5cm}44.74 &	\hspace{-0.5cm}$-$   &	\hspace{-0.5cm}$-$ &	\hspace{-0.05cm}1.99$^{\pm0.14}$ &	$-$\\
J0817$+$2237&	$-$  & 45.24&	\hspace{-0.4cm}$-$&	\hspace{-0.4cm}$-$& \hspace{-0.4cm}41.95 &	\hspace{-0.05cm}$-$&	\hspace{-0.5cm}$-$   &\hspace{-0.5cm}45.21 &	\hspace{-0.5cm}$-$   &	\hspace{-0.5cm}$-$ &	\hspace{-0.05cm}6.72$^{\pm0.95}$ &	$-$\\
J0828$+$3935&	$-$  & 41.57&	\hspace{-0.4cm}$-$&	\hspace{-0.4cm}$-$& \hspace{-0.4cm}15.63 &	\hspace{-0.05cm}$-$&	\hspace{-0.5cm}$-$   &\hspace{-0.5cm}44.61 &	\hspace{-0.5cm}$-$   &	\hspace{-0.5cm}$-$ &	\hspace{-0.05cm}2.85$^{\pm0.54}$ &	$-$\\
J0839$+$1921&	21.60& 36.12&	\hspace{-0.4cm}$-$&	\hspace{-0.4cm}$-$& \hspace{-0.4cm}23.96 &	\hspace{-0.05cm}$-$&	\hspace{-0.5cm}$-$   &\hspace{-0.5cm}45.29 &	\hspace{-0.5cm}$-$   &	\hspace{-0.5cm}$-$ &	\hspace{-0.05cm}4.72$^{\pm0.10}$ &	$-$\\
J0904$+$2819&	$-$  & 36.98&	\hspace{-0.4cm}$-$&	\hspace{-0.4cm}$-$& \hspace{-0.4cm}58.68 &	\hspace{-0.05cm}$-$&	\hspace{-0.5cm}$-$   &\hspace{-0.5cm}45.44 &	\hspace{-0.5cm}$-$   &	\hspace{-0.5cm}$-$ &	\hspace{-0.05cm}5.88$^{\pm0.26}$ &	$-$\\
J0906$+$0832&	38.41& 48.46&	\hspace{-0.4cm}$-$&	\hspace{-0.4cm}$-$& \hspace{-0.4cm}8.01  &	\hspace{-0.05cm}$-$&	\hspace{-0.5cm}$-$   &\hspace{-0.5cm}44.79 &	\hspace{-0.5cm}$-$   &	\hspace{-0.5cm}$-$ &	\hspace{-0.05cm}4.78$^{\pm2.13}$ &	$-$\\
J0924$+$3547&	$-$  & 46.75&	\hspace{-0.4cm}$-$&	\hspace{-0.4cm}$-$& \hspace{-0.4cm}19.14 &	\hspace{-0.05cm}$-$&	\hspace{-0.5cm}$-$   &\hspace{-0.5cm}45.06 &	\hspace{-0.5cm}$-$   &	\hspace{-0.5cm}$-$ &	\hspace{-0.05cm}6.09$^{\pm0.73}$ &	$-$\\
J0925$+$1444&	$-$  & 39.36&	\hspace{-0.4cm}$-$&	\hspace{-0.4cm}$-$& \hspace{-0.4cm}32.33 &	\hspace{-0.05cm}$-$&	\hspace{-0.5cm}$-$   &\hspace{-0.5cm}45.03 &	\hspace{-0.5cm}$-$   &	\hspace{-0.5cm}$-$ &	\hspace{-0.05cm}4.18$^{\pm0.41}$ &	$-$\\
J0935$+$0204&	$-$  & 61.33&	\hspace{-0.4cm}142.40&	\hspace{-0.4cm}$-$& \hspace{-0.4cm}92.04 &	\hspace{-0.05cm}17.1&	\hspace{-0.5cm}$-$   &\hspace{-0.5cm}45.26 &	\hspace{-0.5cm}44.76 &	\hspace{-0.5cm}$-$ &	\hspace{-0.05cm}13.22$^{\pm0.75}$ &	15.07$^{\pm1.44}$\\
J0941$+$3853&	$-$  & 59.18&	\hspace{-0.4cm}234.00&	\hspace{-0.4cm}$-$& \hspace{-0.4cm}42.59 &	\hspace{-0.05cm}9.19&	\hspace{-0.5cm}$-$   &\hspace{-0.5cm}44.89 &	\hspace{-0.5cm}44.45 &	\hspace{-0.5cm}$-$ &	\hspace{-0.05cm}8.01$^{\pm1.03}$ &	28.53$^{\pm2.11}$\\
J0952$+$2352&	$-$  & 36.19&	\hspace{-0.4cm}$-$&	\hspace{-0.4cm}$-$& \hspace{-0.4cm}53.89 &	\hspace{-0.05cm}$-$&	\hspace{-0.5cm}$-$   &\hspace{-0.5cm}45.31 &	\hspace{-0.5cm}$-$   &	\hspace{-0.5cm}$-$ &	\hspace{-0.05cm}4.85$^{\pm0.49}$ &	$-$\\
J1000$+$0005&	$-$  & 30.90&	\hspace{-0.4cm}$-$&	\hspace{-0.4cm}$-$& \hspace{-0.4cm}12.99 &	\hspace{-0.05cm}$-$&	\hspace{-0.5cm}$-$   &\hspace{-0.5cm}44.65 &	\hspace{-0.5cm}$-$   &	\hspace{-0.5cm}$-$ &	\hspace{-0.05cm}1.65$^{\pm0.20}$ &	$-$\\
J1004$+$2225&	$-$  & 44.52&	\hspace{-0.4cm}$-$&	\hspace{-0.4cm}$-$& \hspace{-0.4cm}18.23 &	\hspace{-0.05cm}$-$&	\hspace{-0.5cm}$-$   &\hspace{-0.5cm}44.84 &	\hspace{-0.5cm}$-$   &	\hspace{-0.5cm}$-$ &	\hspace{-0.05cm}4.29$^{\pm0.29}$ &	$-$\\
J1005$+$5019&	18.18& 46.60&	\hspace{-0.4cm}$-$&\hspace{-0.4cm}78.57&\hspace{-0.4cm}9.60  &	\hspace{-0.05cm}$-$&\hspace{-0.5cm}$-$   & \hspace{-0.5cm}44.98 &	\hspace{-0.5cm}$-$   &\hspace{-0.5cm}3.74$^{\pm0.46}$&\hspace{-0.05cm}5.53$^{\pm2.99}$&	$-$\\
J1006$+$3236&	$-$  & 34.69&	\hspace{-0.4cm}$-$&	\hspace{-0.4cm}$-$& \hspace{-0.4cm}8.954 &	\hspace{-0.05cm}$-$&	\hspace{-0.5cm}$-$   &\hspace{-0.5cm}44.57 &	\hspace{-0.5cm}$-$   &	\hspace{-0.5cm}$-$ &	\hspace{-0.05cm}1.90$^{\pm0.20}$ &	$-$\\
J1009$+$0529&	$-$  & 68.01&	\hspace{-0.4cm}$-$&	\hspace{-0.4cm}$-$& \hspace{-0.4cm}71.05 &	\hspace{-0.05cm}$-$&	\hspace{-0.5cm}$-$   &\hspace{-0.5cm}45.41 &	\hspace{-0.5cm}$-$   &	\hspace{-0.5cm}$-$ &	\hspace{-0.05cm}19.25$^{\pm1.07}$&	$-$\\
J1010$+$4132&	$-$  & 28.88&	\hspace{-0.4cm}65.89&	\hspace{-0.4cm}$-$& \hspace{-0.4cm}233.60&	\hspace{-0.05cm}35.86&	\hspace{-0.5cm}$-$   &\hspace{-0.5cm}45.62 &	\hspace{-0.5cm}45.04 &	\hspace{-0.5cm}$-$ &	\hspace{-0.05cm}4.45$^{\pm0.34}$ &	4.45$^{\pm0.76}$\\
J1023$+$6357&	$-$  & 48.48&	\hspace{-0.4cm}$-$&	\hspace{-0.4cm}$-$& \hspace{-0.4cm}53.00 &	\hspace{-0.05cm}$-$&	\hspace{-0.5cm}$-$   &\hspace{-0.5cm}45.43 &	\hspace{-0.5cm}$-$   &	\hspace{-0.5cm}$-$ &	\hspace{-0.05cm}10.05$^{\pm0.97}$ &	$-$\\
J1100$+$1046&	$-$  & 49.35&	\hspace{-0.4cm}21.30&	\hspace{-0.4cm}$-$& \hspace{-0.4cm}60.94 &	\hspace{-0.05cm}11.07&	\hspace{-0.5cm}$-$   &\hspace{-0.5cm}44.76 &	\hspace{-0.5cm}44.25 &	\hspace{-0.5cm}$-$ &	\hspace{-0.05cm}4.79$^{\pm2.17}$ &       0.19$^{\pm0.10}$\\
J1100$+$2314&	$-$  & 66.57&	\hspace{-0.4cm}296.03&	\hspace{-0.4cm}$-$& \hspace{-0.4cm}100.50&	\hspace{-0.05cm}31.02&	\hspace{-0.5cm}$-$   &\hspace{-0.5cm}45.19 &	\hspace{-0.5cm}44.91&	\hspace{-0.5cm}$-$ &	\hspace{-0.05cm}14.34$^{\pm4.02}$ &	77.29$^{\pm5.26}$\\
J1107$+$0547&	33.64& 40.26&	\hspace{-0.4cm}$-$&	\hspace{-0.4cm}$-$& \hspace{-0.4cm}6.66  &	\hspace{-0.05cm}$-$&	\hspace{-0.5cm}$-$   &\hspace{-0.5cm}44.76 &	\hspace{-0.5cm}$-$   &	\hspace{-0.5cm}$-$ &	\hspace{-0.05cm}3.21$^{\pm1.52}$ &	$-$\\
J1107$+$1628&	$-$  & 36.98&	\hspace{-0.4cm}88.07&	\hspace{-0.4cm}$-$& \hspace{-0.4cm}180.00&	\hspace{-0.05cm}33.76&	\hspace{-0.5cm}$-$   &\hspace{-0.5cm}45.53 &	\hspace{-0.5cm}45.04 &	\hspace{-0.5cm}$-$ &	\hspace{-0.05cm}6.57$^{\pm0.39}$ &	7.92$^{\pm0.86}$\\
J1110$+$0321&	$-$  & 29.41&	\hspace{-0.4cm}$-$&	\hspace{-0.4cm}$-$& \hspace{-0.4cm}16.59 &	\hspace{-0.05cm}$-$&	\hspace{-0.5cm}$-$   &\hspace{-0.5cm}44.79 &	\hspace{-0.5cm}$-$   &	\hspace{-0.5cm}$-$ &	\hspace{-0.05cm}1.77$^{\pm0.74}$ &	$-$\\
J1118$+$3828&	$-$  & 39.03&	\hspace{-0.4cm}431.77&	\hspace{-0.4cm}$-$& \hspace{-0.4cm}29.22 &	\hspace{-0.05cm}3.46&	\hspace{-0.5cm}$-$   &\hspace{-0.5cm}44.87 &	\hspace{-0.5cm}44.17 &	\hspace{-0.5cm}$-$ &	\hspace{-0.05cm}3.39$^{\pm1.22}$ &	69.97$^{\pm29.70}$\\
J1119$+$3858&	$-$  & 64.70&	\hspace{-0.4cm}345.40&	\hspace{-0.4cm}$-$& \hspace{-0.4cm}31.08 &	\hspace{-0.05cm}6.27&	\hspace{-0.5cm}$-$   &\hspace{-0.5cm}44.88 &	\hspace{-0.5cm}44.41 &	\hspace{-0.5cm}$-$ &	\hspace{-0.05cm}9.47$^{\pm3.54}$ &	59.42$^{\pm20.60}$\\
J1158$+$6254&	$-$  & 66.27&	\hspace{-0.4cm}284.30&	\hspace{-0.4cm}$-$& \hspace{-0.4cm}185.80&	\hspace{-0.05cm}44.74&	\hspace{-0.5cm}$-$   &\hspace{-0.5cm}45.50 &	\hspace{-0.5cm}45.11 &	\hspace{-0.5cm}$-$ &	\hspace{-0.05cm}20.31$^{\pm2.35}$&	89.96$^{\pm19.39}$\\
J1217$+$1019&	20.56& 38.45&	\hspace{-0.4cm}$-$&\hspace{-0.4cm}59.73&\hspace{-0.4cm}5.42  &	\hspace{-0.05cm}$-$&\hspace{-0.5cm}45.39 & \hspace{-0.5cm}44.70 &	\hspace{-0.5cm}$-$   &\hspace{-0.5cm}3.96$^{\pm0.53}$&\hspace{-0.05cm}2.71$^{\pm1.00}$&	$-$\\
J1223$+$3707&	$-$  & 50.07&	\hspace{-0.4cm}264.00&	\hspace{-0.4cm}$-$& \hspace{-0.4cm}34.43 &	\hspace{-0.05cm}9.27&	\hspace{-0.5cm}$-$   &\hspace{-0.5cm}44.63 &	\hspace{-0.5cm}44.29 &	\hspace{-0.5cm}$-$ &	\hspace{-0.05cm}4.24$^{\pm0.95}$ &	30.00$^{\pm5.01}$\\
J1236$+$1034&	$-$  & 38.83&	\hspace{-0.4cm}272.90&	\hspace{-0.4cm}$-$& \hspace{-0.4cm}53.47 &	\hspace{-0.05cm}11.31&	\hspace{-0.5cm}$-$   &\hspace{-0.5cm}45.05 &	\hspace{-0.5cm}44.60 &	\hspace{-0.5cm}$-$ &	\hspace{-0.05cm}4.13$^{\pm1.94}$ &	46.03$^{\pm9.20}$\\
J1256$+$1008&	$-$  & 33.46&	\hspace{-0.4cm}33.74&	\hspace{-0.4cm}$-$& \hspace{-0.4cm}15.93 &	\hspace{-0.05cm}$-$&	\hspace{-0.5cm}$-$   &\hspace{-0.5cm}44.67 &	\hspace{-0.5cm}$-$   &	\hspace{-0.5cm}$-$ &	\hspace{-0.05cm}1.99$^{\pm0.18}$ &	$-$\\
J1319$+$5148&	$-$  & 39.67&	\hspace{-0.4cm}$-$&	\hspace{-0.4cm}$-$& \hspace{-0.4cm}76.29 &	\hspace{-0.05cm}$-$&	\hspace{-0.5cm}$-$   &\hspace{-0.5cm}45.52 &	\hspace{-0.5cm}$-$   &	\hspace{-0.5cm}$-$ &	\hspace{-0.05cm}7.42$^{\pm0.80}$ &	$-$\\
J1334$+$5501&	$-$  & 55.70&	\hspace{-0.4cm}$-$&	\hspace{-0.4cm}$-$& \hspace{-0.4cm}21.12 &	\hspace{-0.05cm}$-$&	\hspace{-0.5cm}$-$   &\hspace{-0.5cm}45.06 &	\hspace{-0.5cm}$-$   &	\hspace{-0.5cm}$-$ &	\hspace{-0.05cm}8.63$^{\pm0.92}$ &	$-$\\
J1358$+$5752&	$-$  & 45.54&	\hspace{-0.4cm}$-$&	\hspace{-0.4cm}$-$& \hspace{-0.4cm}67.78 &	\hspace{-0.05cm}$-$&	\hspace{-0.5cm}$-$   &\hspace{-0.5cm}45.62 &	\hspace{-0.5cm}$-$   &	\hspace{-0.5cm}$-$ &	\hspace{-0.05cm}11.05$^{\pm1.92}$ &	$-$\\
J1425$+$2404&	$-$  & 56.62&	\hspace{-0.4cm}131.60&	\hspace{-0.4cm}$-$& \hspace{-0.4cm}89.40 &	\hspace{-0.05cm}19.66&	\hspace{-0.5cm}$-$   &\hspace{-0.5cm}45.25 &	\hspace{-0.5cm}44.83 &	\hspace{-0.5cm}$-$ &	\hspace{-0.05cm}11.16$^{\pm1.62}$ &	13.87$^{\pm1.64}$\\
J1433$+$3209&	$-$  & 45.42&	\hspace{-0.4cm}$-$&	\hspace{-0.4cm}$-$& \hspace{-0.4cm}2.924 &	\hspace{-0.05cm}$-$&	\hspace{-0.5cm}$-$   &\hspace{-0.5cm}44.02 &	\hspace{-0.5cm}$-$   &	\hspace{-0.5cm}$-$ &	\hspace{-0.05cm}1.73$^{\pm1.85}$ &	$-$\\
J1513$+$1011&	23.23& 42.70&	\hspace{-0.4cm}$-$&	\hspace{-0.4cm}$-$& \hspace{-0.4cm}36.67 &	\hspace{-0.05cm}$-$&	\hspace{-0.5cm}$-$   &\hspace{-0.5cm}45.42 &	\hspace{-0.5cm}$-$   &	\hspace{-0.5cm}$-$ &	\hspace{-0.05cm}7.72$^{\pm0.57}$ &	$-$\\
J1550$+$3652&	21.99& 47.58&	\hspace{-0.4cm}$-$&\hspace{-0.4cm}41.78&\hspace{-0.4cm}4.80  &	\hspace{-0.05cm}$-$&\hspace{-0.5cm}45.28 & \hspace{-0.5cm}44.69 &	\hspace{-0.5cm}$-$   &\hspace{-0.5cm}3.97$^{\pm0.52}$&\hspace{-0.05cm}4.12$^{\pm1.81}$&	$-$\\
J1557$+$0253&	10.42& 24.83&	\hspace{-0.4cm}$-$&\hspace{-0.4cm}34.97&\hspace{-0.4cm}4.69  &	\hspace{-0.05cm}$-$&\hspace{-0.5cm}45.19 & \hspace{-0.5cm}44.66 &	\hspace{-0.5cm}$-$   &\hspace{-0.5cm}0.79$^{\pm0.16}$&\hspace{-0.05cm}1.09$^{\pm1.28}$&	$-$\\
J1557$+$3304&	$-$  & 52.94&	\hspace{-0.4cm}$-$&	\hspace{-0.4cm}$-$& \hspace{-0.4cm}25.46 &	\hspace{-0.05cm}$-$&	\hspace{-0.5cm}$-$   &\hspace{-0.5cm}44.97 &	\hspace{-0.5cm}$-$   &	\hspace{-0.5cm}$-$ &	\hspace{-0.05cm}7.05$^{\pm1.15}$ &	$-$\\
J1622$+$3531&	$-$  & 43.02&	\hspace{-0.4cm}$-$&	\hspace{-0.4cm}$-$& \hspace{-0.4cm}10.56 &	\hspace{-0.05cm}$-$&	\hspace{-0.5cm}$-$   &\hspace{-0.5cm}44.86 &	\hspace{-0.5cm}$-$   &	\hspace{-0.5cm}$-$ &	\hspace{-0.05cm}4.08$^{\pm0.83}$ &	$-$\\					
J1623$+$3419&	25.17& 48.20&	\hspace{-0.4cm}$-$&\hspace{-0.4cm}14.37&\hspace{-0.4cm}2.59  &	\hspace{-0.05cm}$-$&\hspace{-0.5cm}44.80 & \hspace{-0.5cm}44.40 &	\hspace{-0.5cm}$-$   &\hspace{-0.5cm}2.88$^{\pm0.59}$&\hspace{-0.05cm}3.04$^{\pm1.78}$&	$-$\\
J2335$-$0927&	14.18& 35.59&	\hspace{-0.4cm}$-$&\hspace{-0.4cm}60.11&\hspace{-0.4cm}1.14  &	\hspace{-0.05cm}$-$&\hspace{-0.5cm}45.38 & \hspace{-0.5cm}44.00 &	\hspace{-0.5cm}$-$   &\hspace{-0.5cm}1.85$^{\pm0.26}$&\hspace{-0.05cm}1.04$^{\pm0.99}$&	$-$\\
\hline
\end{tabular}
\end{minipage}
\end{table*}							

\begin{table*}
 \centering
 \begin{minipage}{140mm}
  \caption{Optical luminosity and accretion rate for GRQs.}
  \begin{tabular}{@{}lccccccccc@{}}
\hline
IAU	      &	&log(L$_{bol})$ & 	&  & log(L$_{Edd}$) &  &\.m1350 &\.m3000 & \.m5100\\								
name	      &1350\AA	        &3000\AA        & 5100\AA	        & CIV           & MgII           & $H\beta$       &        &         &        \\							
              & \multicolumn{3}{c|}{\rule[0mm]{4cm}{0.1pt}}             &\multicolumn{3}{c|}{\rule[0mm]{4cm}{0.1pt}}               &        &         &       \\
              &\multicolumn{3}{c}{erg s$^{-1}$}               &\multicolumn{3}{c}{erg s$^{-1}$}       &        &         &        \\
\hline
J0204$-$0944	&$-$	&45.38	&$-$	&$-$	&46.45	&$-$	&$-$	&0.09 &$-$\\
J0210$+$0118	&$-$	&45.94	&$-$	&$-$	&47.05	&$-$	&$-$	&0.08 &$-$\\
J0754$+$3033	&$-$	&45.94	&45.66	&$-$	&47.04	&47.29	&$-$	&0.08 &0.02\\
J0754$+$4316	&$-$	&$-$	&45.48	&$-$	&$-$	&47.63	&$-$	&$-$ &0.01\\
J0801$+$4736	&$-$	&$-$	&43.92	&$-$	&$-$	&46.33	&$-$	&$-$&0.004\\
J0809$+$2912	&$-$	&46.25	&$-$	&$-$	&47.16	&$-$	&$-$	&0.12&$-$\\
J0812$+$3031	&$-$	&45.49	&$-$	&$-$	&46.42	&$-$	&$-$	&0.12&$-$\\
J0819$+$0549	&45.29	&44.86	&$-$	&47.49	&46.66	&$-$	&$-$	&0.02&$-$\\
J0842$+$2147	&$-$	&45.51	&$-$	&$-$	&46.49	&$-$	&$-$	&0.11&$-$\\
J0902$+$5707	&45.97	&45.67	&$-$	&47.08	&46.88	&$-$	&$-$	&0.06&$-$\\
J0918$+$2325	&$-$	&45.85	&45.48	&$-$	&47.11	&47.39	&$-$	&0.06&0.01\\
J0925$+$4004	&$-$	&45.87	&45.55	&$-$	&47.21	&47.54	&$-$	&0.05&0.01\\
J0937$+$2937	&$-$	&45.69	&45.38	&$-$	&46.77	&46.85	&$-$	&0.08&0.03\\
J0944$+$2331	&$-$	&45.90	&$-$	&$-$	&46.92	&$-$	&$-$	&0.10&$-$\\
J0959$+$1216	&$-$	&45.58	&$-$	&$-$	&46.82	&$-$	&$-$	&0.06&$-$\\
J1020$+$0447	&$-$	&45.26	&$-$	&$-$	&46.83	&$-$	&$-$	&0.03&$-$\\
J1020$+$3958	&$-$	&45.77	&$-$	&$-$	&47.22	&$-$	&$-$	&0.04&$-$\\
J1030$+$5310	&$-$	&45.90	&$-$	&$-$	&46.76	&$-$	&$-$	&0.14&$-$\\
J1054$+$4152	&$-$	&45.78	&$-$	&$-$	&46.96	&$-$	&$-$	&0.07&$-$\\
J1056$+$4100	&45.34	&45.16	&$-$	&46.83	&46.70	&$-$	&$-$	&0.03&$-$\\
J1145$-$0033	&45.53	&45.51	&$-$	&47.43	&$-$	&$-$	&0.01	&$-$&$-$\\
J1151$+$3355	&$-$	&45.73	&$-$	&$-$	&46.97	&$-$	&$-$	&0.06&$-$\\
J1229$+$3555	&$-$	&45.37	&$-$	&$-$	&46.39	&$-$	&$-$	&0.10&$-$\\
J1304$+$2454	&$-$	&45.92	&$-$	&$-$	&47.01	&47.57	&$-$	&0.08&$-$\\
J1321$+$3741	&$-$	&45.64	&$-$	&$-$	&47.25	&$-$	&$-$	&0.03&$-$\\
J1340$+$4232	&$-$	&45.47	&$-$	&$-$	&46.87	&$-$	&$-$	&0.04&$-$\\
J1353$+$2631    &$-$	&45.63	&45.46	&$-$	&46.75	&47.54	&$-$	&0.08&0.01\\
J1410$+$2955	&$-$	&45.65	&$-$	&$-$	&47.20	&$-$	&$-$	&0.03&$-$\\
J1427$+$2632    &$-$	&$-$	&45.67	&$-$	&$-$	&47.59	&$-$	&$-$&0.01\\
J1432$+$1548    &$-$	&45.65	&$-$	&$-$	&46.96	&$-$	&$-$	&0.05&$-$\\
J1723$+$3417    &$-$	&45.39	&45.73	&$-$	&$-$	&46.66	&$-$	&$-$&0.12\\
J2344$-$0032	&$-$	&45.723	&$-$	&$-$	&46.78	&$-$	&$-$	&0.09&$-$\\					
\hline
\end{tabular}
\end{minipage}\\
\end{table*}

\begin{table*}
 \centering
 \begin{minipage}{140mm}
  \caption{Optical luminosity and accretion rate for small-size radio quasars.}
  \begin{tabular}{@{}lccccccccc@{}}
\hline
IAU	      &	&log(L$_{bol})$ & 	&  & log(L$_{Edd}$) &  &\.m1350 &\.m3000 & \.m5100\\								
name	      &1350\AA	        &3000\AA        & 5100\AA	        & CIV           & MgII           & $H\beta$       &        &         &        \\	
						              & \multicolumn{3}{c|}{\rule[0mm]{4cm}{0.1pt}}             &\multicolumn{3}{c|}{\rule[0mm]{4cm}{0.1pt}}               &        &         &       \\
              &\multicolumn{3}{c|}{erg s$^{-1}$}   & \multicolumn{3}{c|}{erg s$^{-1}$}       &        &         &        \\  
\hline
J0034$+$0118	&$-$	&45.35	&$-$	&$-$	&46.87	&$-$	&$-$	&0.03 &$-$\\
J0051$-$0902	&$-$	&45.58	&$-$	&$-$	&47.10	&$-$	&$-$	&0.03 &$-$\\
J0130$-$0135	&$-$	&45.83	&$-$	&$-$	&47.19	&$-$	&$-$	&0.05 &$-$\\
J0245$+$0108	&46.05	&45.73	&$-$	&47.23	&47.13	&$-$	&0.07	&0.04 &$-$\\
J0745$+$3142	&$-$	&46.51	&46.26	&$-$	&47.41	&47.78	&$-$	&0.13 &0.03\\
J0811$+$2845	&46.06	&45.85	&$-$	&47.00	&47.08	&$-$	&0.11	&0.06 &$-$\\
J0814$+$3237	&$-$	&45.51	&$-$	&$-$	&46.46	&45.89	&$-$	&0.11 &$-$\\
J0817$+$2237	&$-$	&45.98	&$-$	&$-$	&46.99	&$-$	&$-$	&0.10 &$-$\\
J0828$+$3935	&$-$	&45.38	&$-$	&$-$	&46.62	&$-$	&$-$	&0.06 &$-$\\
J0839$+$1921	&46.33	&46.06	&$-$	&46.95	&46.84	&$-$	&0.24	&0.17 &$-$\\
J0904$+$2819	&$-$	&46.21	&$-$	&$-$	&46.93	&$-$	&$-$	&0.19 &$-$\\
J0906$+$0832	&45.82	&45.56	&$-$	&47.18	&46.84	&$-$	&0.04	&0.05 &$-$\\
J0924$+$3547	&$-$	&45.83	&$-$	&$-$	&46.95	&$-$	&$-$	&0.08 &$-$\\
J0925$+$1444	&$-$	&45.81	&$-$	&$-$	&46.78	&$-$	&$-$	&0.11 &$-$\\
J0935$+$0204	&$-$	&46.03	&45.72	&$-$	&47.28	&47.34	&$-$	&0.06 &0.02\\
J0941$+$3853	&$-$	&45.66	&45.41	&$-$	&47.07	&47.62	&$-$	&0.04 &0.01\\
J0952$+$2352	&$-$	&46.080	&$-$	&$-$	&46.85	&$-$	&$-$	&0.17 &$-$\\
J1000$+$0005	&$-$	&45.42	&$-$	&$-$	&46.38	&$-$	&$-$	&0.11 &$-$\\
J1004$+$2225	&$-$	&45.61	&$-$	&$-$	&46.79	&$-$	&$-$	&0.07 &$-$\\
J1005$+$5019	&46.21	&45.75	&$-$	&46.74	&46.90	&$-$	&0.30	&0.07 &$-$\\
J1006$+$3236	&$-$	&45.34	&$-$	&$-$	&46.44	&$-$	&$-$	&0.08 &$-$\\
J1009$+$0529	&$-$	&46.18	&$-$	&$-$	&47.45	&$-$	&$-$	&0.05 &$-$\\
J1010$+$4132	&$-$	&46.40	&46.00	&$-$	&46.81	&46.81	&$-$	&0.39 &0.15\\
J1023$+$6357	&$-$	&46.20	&$-$	&$-$	&47.16	&$-$	&$-$	&0.11 &$-$\\
J1100$+$1046	&$-$	&45.53	&45.20	&$-$	&46.84	&45.43	&$-$	&0.05 &0.59\\
J1100$+$2314	&$-$	&45.96	&45.87	&$-$	&47.32	&48.05	&$-$	&0.04 &0.01\\
J1107$+$0547	&45.87	&45.54	&$-$	&47.09	&46.67	&$-$	&0.06	&0.07 &$-$\\
J1107$+$1628	&$-$	&46.31	&45.99	&$-$	&46.98	&47.06	&$-$	&0.21 &0.09\\
J1110$+$0321	&$-$	&45.57	&$-$	&$-$	&46.41	&$-$	&$-$	&0.14 &$-$\\
J1118$+$3828	&$-$	&45.64	&45.12	&$-$	&46.69	&48.01	&$-$	&0.09 &0.001\\
J1119$+$3858	&$-$	&45.65	&45.37	&$-$	&47.14	&47.94	&$-$	&0.03 &0.003\\
J1158$+$6254	&$-$	&46.27	&46.07	&$-$	&47.47	&48.12	&$-$	&0.06 &0.01\\
J1217$+$1019	&46.06	&45.47	&$-$	&46.76	&46.60	&$-$	&0.20	&0.08 &$-$\\
J1223$+$3707	&$-$	&45.40	&45.24	&$-$	&46.79	&47.64	&$-$	&0.04 &0.004\\
J1236$+$1034	&$-$	&45.82	&45.56	&$-$	&46.78	&47.82	&$-$	&0.11 &0.01\\
J1256$+$1008	&$-$	&45.44	&$-$	&$-$	&46.46	&45.70	&$-$	&0.10 &$-$\\
J1319$+$5148	&$-$	&46.29	&$-$	&$-$	&47.03	&$-$	&$-$	&0.18 &$-$\\
J1334$+$5501	&$-$	&45.83	&$-$	&$-$	&47.10	&$-$	&$-$	&0.05 &$-$\\
J1358$+$5752	&$-$	&46.39	&$-$	&$-$	&47.20	&$-$	&$-$	&0.16 &$-$\\
J1425$+$2404	&$-$	&46.03	&45.78	&$-$	&47.21	&47.30	&$-$	&0.07 &0.03\\
J1433$+$3209	&$-$	&44.79	&$-$	&$-$	&46.40	&$-$	&$-$	&0.03 &$-$\\
J1513$+$1011	&46.59	&46.20	&$-$	&47.15	&47.05	&$-$	&0.27	&0.14 &$-$\\
J1550$+$3652	&45.95	&45.46	&$-$	&46.76	&46.78	&$-$	&0.15	&0.05 &$-$\\
J1557$+$0253	&45.85	&45.43	&$-$	&46.06	&46.20	&$-$	&0.61	&0.17 &$-$\\
J1557$+$3304	&$-$	&45.74	&$-$	&$-$	&47.01	&$-$	&$-$	&0.05 &$-$\\
J1622$+$3531	&$-$	&45.63	&$-$	&$-$	&46.77	&$-$	&$-$	&0.07 &$-$\\
J1623$+$3419	&45.46	&45.17	&$-$	&46.62	&46.64	&$-$	&0.07	&0.03 &$-$\\
J2335$-$0927	&46.04	&44.77	&$-$	&46.43	&46.18	&$-$	&0.41	&0.04 &$-$\\
\hline
\end{tabular}
\end{minipage}\\
\end{table*}

\label{lastpage}

\end{document}